\documentclass{sig-alternate} %[conference]{IEEEtran}

\newcommand{\lara}{\textsc{Lara}}%
\newcommand{\laradb}{\textsc{LaraDB}}%
\newcommand{\plara}{\textsc{PLara}}%

\usepackage[dvipsnames,table]{xcolor}
\usepackage{bm}
\usepackage{amsmath}
\usepackage{amssymb}
\usepackage{amscd}
\usepackage{mathpartir}

\usepackage{subfig}
\usepackage{adjustbox}
\usepackage{calc}
\usepackage{color}
\usepackage{varwidth}
\usepackage{multirow}

\newlength{\centersep}
\usepackage{relsize}

\newcommand{\newcomment}[3]{%
  \expandafter\newcommand\csname #1\endcsname[1]{{\color{#2}{[#3: ##1]}}}%
}
\newcomment{bill}{blue}{Bill}

\usepackage{ wasysym }

\newcommand\uproduct{%
  \mathchoice{\mathbin{\;\rotatebox{90}{$\Bowtie$}}}%
             {\mathbin{\;\rotatebox{90}{$\Bowtie$}}}%
             {\mathbin{\;\rotatebox{90}{\scalebox{0.65}{$\Bowtie$}}}}%
             {\mathbin{\;\rotatebox{90}{\scalebox{0.65}{$\Bowtie$}}}}%
}

\newcommand{\belse}{\text{ else }}

\newcommand{\temperature}{temp}
\newcommand{\humidity}{hum}

\newcommand{\overbar}[1]{\mkern 1.5mu\overline{\mkern-1.5mu#1\mkern-1.5mu}\mkern 1.5mu}
\newcommand{\tup}[1]{\overbar{#1}}
\newcommand{\ex}[0]{\mathbb{E}}

\newcommand{\bb}[1]{\textsc{#1}}
\newcommand{\bsort}[0]{\textbf{S{\scriptsize ORT}}}
\newcommand{\bext}{\bb{Ext}}
\newcommand{\bmap}{\bb{Map}}
\newcommand{\bby}{\bb{by}}
\newcommand{\bon}{\bb{on}}
\newcommand{\bto}{\bb{to}}
\newcommand{\bfrom}{\bb{from}}
\newcommand{\bjoin}{\bb{Join}}
\newcommand{\bunion}{\bb{Union}}
\newcommand{\bagg}{\bb{Agg}}
\newcommand{\bselect}{\bb{select}}
\newcommand{\bas}{\bb{as}}
\newcommand{\bgroupby}{\bb{group by}}
\newcommand{\brename}{\bb{Rename}}
\newcommand{\bmergejoin}{\bb{MergeJoin}}
\newcommand{\bmergeunion}{\bb{MergeUnion}}
\newcommand{\bmergeagg}{\bb{MergeAgg}}
\newcommand{\bsortagg}{\textbf{S{\scriptsize ORT}A{\scriptsize GG}}}
\newcommand{\bapply}{\bb{Map}}
\newcommand{\bapplyext}{\bb{Ext}}
\newcommand{\bwhere}{\bb{where}}
\newcommand{\bload}{\bb{Load}}
\newcommand{\bstore}{\bb{Store}}
\newcommand{\bover}{\bb{over}}
\newcommand{\bnull}{\bb{null}}

\usepackage{xparse}
\DeclareDocumentCommand{\join}{ O{} O{} }{\operatorname{\bowtie_{#1}^{#2}}\!}
\newcommand{\joino}[1][]{\join[\otimes][#1]}
\newcommand{\joinp}[1][]{\join[\otimes][#1]} %\odot

\DeclareDocumentCommand{\sjoin}{ O{} O{} }{\operatorname{\hat{\bowtie}_{#1^{#2}}}}

\DeclareDocumentCommand{\union}{ O{} O{} }{\operatorname{\uproduct_{#1}^{#2}}\!}
\newcommand{\uniono}[1][]{\union[\oplus][#1]} %\operatorname{\uproduct_\oplus^{#1}}
\newcommand{\unionp}[1][]{\union[+][#1]} %\operatorname{\uproduct_\oplus^{#1}}
\newcommand{\tupoplus}[1][]{\operatorname{\tup{\oplus}}^{#1}}
\DeclareMathOperator{\ext}{ext}

\DeclareMathOperator{\map}{map}

\DeclareMathOperator{\trace}{tr}

\usepackage{mathtools}

\newcommand{\tr}[0]{{\intercal}}

\newcommand{\mxm}{{\,\oplus.\otimes\,}}
\makeatletter
\newcommand{\removelatexerror}{\let\@latex@error\@gobble}
\makeatother
\newlength{\algspace}
\usepackage[]{algorithm2e}

\usepackage[  
  bookmarks=false,
  hyperfootnotes=false,
  hyperindex=false,
  hidelinks,   
    pdftitle={LaraDB: A Minimalist Kernel for Linear and Relational Algebra Computation},  
  pdfauthor={Dylan Hutchison, Bill Howe, Dan Suciu},
  pdfsubject={Database query operators},
  pdfpagelabels=false]{hyperref}

\newlength\myindent
\usepackage{todonotes}

\graphicspath{{.}{./figures/}}

\usepackage{siunitx}
\usepackage{threeparttable}
\usepackage{multirow}
\usepackage{adjustbox}
\usepackage{array}
\newcolumntype{R}[2]{%
    >{\adjustbox{angle=#1,lap=\width-(#2)}\bgroup}%
    l%
    <{\egroup}%
}

\usepackage{listings}

\begin{document}

\begin{CCSXML}
<ccs2012>
<concept>
<concept_id>10002951.10002952.10003190.10003192.10003398</concept_id>
<concept_desc>Information systems~Query operators</concept_desc>
<concept_significance>500</concept_significance>
</concept>
<concept>
<concept_id>10002951.10002952.10003197</concept_id>
<concept_desc>Information systems~Query languages</concept_desc>
<concept_significance>300</concept_significance>
</concept>
<concept>
<concept_id>10002951.10002952.10003190.10003192.10003210</concept_id>
<concept_desc>Information systems~Query optimization</concept_desc>
<concept_significance>100</concept_significance>
</concept>
<concept>
<concept_id>10002951.10002952.10002953.10010819</concept_id>
<concept_desc>Information systems~Physical data models</concept_desc>
<concept_significance>100</concept_significance>
</concept>
</ccs2012>
\end{CCSXML}

\ccsdesc[500]{Information systems~Query operators}
\ccsdesc[300]{Information systems~Query languages}
\ccsdesc[100]{Information systems~Query optimization}
\ccsdesc[100]{Information systems~Physical data models}

% Original::::
% \copyrightyear{2017} 
% \acmYear{2017} 
% \setcopyright{acmcopyright}
% \acmConference{BeyondMR'17}{May 19, 2017}{Chicago, IL, USA}
% \acmPrice{15.00}
% \acmDOI{http://dx.doi.org/10.1145/3070607.3070608}
% \acmISBN{978-1-4503-5019-8/17/05}

% Copyright
\setcopyright{acmcopyright}
%\setcopyright{acmlicensed}
% \setcopyright{rightsretained}
%\setcopyright{usgov}
%\setcopyright{usgovmixed}
%\setcopyright{cagov}
%\setcopyright{cagovmixed}

% DOI
\doi{http://dx.doi.org/10.1145/3070607.3070608}

% ISBN
\isbn{978-1-4503-5019-8/17/05}

%Conference
\conferenceinfo{BeyondMR'17}{May 19, 2017, Chicago, IL, USA}
\acmPrice{\$15.00}

%
% --- Author Metadata here ---
\CopyrightYear{2017} % Allows default copyright year (20XX) to be over-ridden - IF NEED BE.
% \crdata{978-1-4503-5019-8/17/05}  % Allows default copyright data (0-89791-88-6/97/05) to be over-ridden - IF NEED BE.
% --- End of Author Metadata ---

\title{LaraDB: A Minimalist Kernel for \\ Linear and Relational Algebra Computation}
% \subtitle{[Extended Abstract]
% \titlenote{A full version of this paper is available as
% \textit{Author's Guide to Preparing ACM SIG Proceedings Using
% \LaTeX$2_\epsilon$\ and BibTeX} at
% \texttt{www.acm.org/eaddress.htm}}}

\numberofauthors{1}
\author{
\alignauthor
	Dylan Hutchison, Bill Howe, Dan Suciu\\[3pt]
     \affaddr{\{dhutchis,billhowe,suciu\}@cs.washington.edu}%
       % \affaddr{Department of Computer Science and Engineering}\\
       % \affaddr{University of Washington}\\
       % \affaddr{Seattle, WA 98195-2350, U.S.A.}
% \begin{tabular}{c}
% %\alignauthor
% 	David Maier\\
% 	\affaddr{maier@cs.washington.edu}\\
%        \affaddr{Department of Computer Science}\\
%        \affaddr{Portland State University}\\
%        \affaddr{Portland, OR  97207-0751, U.S.A.}
% \end{tabular}
% 
% \vspace{-20em}%
}

% RA and LA need to be used together
% 

\maketitle
\begin{abstract}
Analytics tasks manipulate structured data with variants of relational algebra (RA) and quantitative data with variants of linear algebra (LA).  
The two computational models have overlapping expressiveness, 
motivating a common programming model that affords unified reasoning and algorithm design.  
At the logical level we propose \lara{}, a lean algebra of three operators, 
that expresses RA and LA as well as relevant optimization rules.  
We show a series of proofs that position \lara{} %formal and informal
at just the right level of expressiveness for a middleware algebra:
more explicit than MapReduce but more general than RA or LA.
At the physical level we find that the \lara{} operators afford efficient implementations using a single primitive that is available in a variety of backend engines: range scans over partitioned sorted maps.

To evaluate these ideas, we implemented the \lara{} operators as range iterators in  Apache Accumulo, a popular implementation of Google's BigTable.  
First we show how \lara{} expresses a sensor quality control task,
and we measure the performance impact of optimizations \lara{} admits on this task.
Second we show that the \laradb{} implementation
outperforms Accumulo's native MapReduce integration 
on a core task involving join and aggregation in the form of matrix multiply,
especially at smaller scales that are typically a poor fit for scale-out approaches.  
We find that \laradb{} offers a conceptually lean framework for optimizing mixed-abstraction analytics tasks, without giving up fast record-level updates and scans.
\end{abstract}
% \printccsdesc

% Some point about how this is more explicit than MapReduce (and, dare we say, lambda calculus)
% while less explicit than either RA or LA. It's at just the right level of expressiveness.

% NoSQL key-value databases emphasize record-level read-write operations, 
% relying on external infrastructure such as Spark or MapReduce 
% to implement complex analytics (e.g., matrix math, relational joins, machine learning). 
% Computing in an external system is expensive for small yet complex analytics, 
% requiring long code paths to extract data from the database and prepare native data structures.
% In response, developers implement custom applications 
% that push simple filters and sums into the database's scans
% in order to maximize in-database processing.
% Recent software generalized this approach 
% to provide native, in-database support for complex analytics.

% In this work we evaluate the performance of in-database vs. external system 
% approaches to query processing for the Apache Accumulo NoSQL database.
% Specifically we run Graphulo, a library for matrix math 
% inside Accumulo's scan-time iterators, and MapReduce, 
% an off-the-shelf external system commonly used with Accumulo,
% on sparse matrix multiplication.
% Results indicate that the Graphulo's in-database approach is superior at smaller problem sizes,
% while at larger problem sizes the two approaches have similar performance.
\section{Introduction}

% Graphulo capitalizes on the idea to do multi-table computation inside the iterators for matrix math
% LaraDB focuses on this new abstraction.

% \begin{enumerate}
% \item ``Implement these three primitives''
% \end{enumerate}

%\begin{figure}[tb]
%\centering
%\includegraphics[]{lara-arch}
%\caption{Architecture of \lara{} as a middleware algebra.
%Front-end languages and APIs translate to \lara{} expressions;
%\lara{} e
%}
%\label{fLaraArch}
%\end{figure}

Analytics tasks involve preprocessing (ETL, restructuring, cleaning) that are typically expressed using relational algebra-based languages as well as numerical tasks (machine learning, optimization, signal processing) that are typically expressed using linear algebra operations \cite{gadepally2014big}.  The distinction between these two programming styles is increasingly blurred: machine learning applications are implemented using RA-oriented interfaces using, for example, Spark, sometimes with extensions for special cases \cite{maas2015gaussian,hellerstein2012madlib,meng2016mllib,bu:12}. 

%\dylan{From previous Lara paper: Pre-processing pipelines are naturally ex-
%pressed in dataflow APIs (e.g., MapReduce, Flink, etc.),
%while machine learning is expressed in linear algebra with
%iterations.}
The relevant tasks involve both programming styles, but systems tend to favor one or the other.  An LA-oriented system may emphasize matrix operations in the programming interface but require awkward gymnastics with column and row indices to implement even simple SPJ queries.  A relational system, in contrast, obscures matrix properties suitable for optimization and algorithm selection.  Some systems allow explicit transformation of relations into matrices and vice versa, exposing a different set of programming idioms for each data type \cite{kunft2016bridging}.  A common programming environment where both styles can be used interchangeably is desirable, as is being explored by a number of systems and libraries, including 
Myria \cite{wang2017myria}, 
Spark \cite{zaharia2010spark}, 
and more.

These systems emphasize mixed-programming syntax but assume more conventional internal computational models, many based on RA.  The benefit of this approach is that conventional RA properties and rewrites are easy to exploit for optimization.  The problem with this approach is that LA properties and rewrites are obscured, or entirely inexpressible.  We find it desirable to use theorems from both LA and RA when reasoning about queries.  For example, the fact that the inner product $U^\tr U$ is symmetric suggests an immediate optimization: only produce its upper triangle rather than also computing its lower triangle redundantly. Although this optimization is possible to implement (and prove) using RA, it is far from obvious, and no RDBMS applies this optimization.  We seek a new set of abstractions to facilitate the use of similar theorem-oriented optimizations.

%Desirable to support systems designed for LA and systems designed for RA in a polystore.  We won't be talking about this in this paper, but we are exploring this in the context of a complex ecosystem of systems with different capabilities.
%\item LA and RA both consist of operators on collections of data.
%Yet they have independent histories, theorems, and implementations.\bill{not sure this point is important.}

We propose a lean algebra of three operators, \lara{}, that subsumes the operators and rules of both LA and RA.  We keep the number of operators to a minimum to simplify an initial implementation on a number of backend systems, and to simplify reasoning during optimization by avoiding large numbers of special cases for relatively simple concepts (pushing selections, etc.).  As with other systems for big data processing \cite{crotty2015tupleware,flume,alexandrov2014stratosphere,chaiken2008scope}, \lara{} operators are parameterized by rich user-defined functions, and the properties of these functions are involved in optimization.   However, \lara{} emphasizes a more restricted semiring structure to capture the properties of vector space algebra as opposed to emphasizing the ``free-for-all'' UDF approach other systems emphasize.

We also propose a physical algebra, \plara{}, that allows reasoning about low-level optimizations such as operator fusion, elimination of unnecessary writes, and shared scans.

Finally we show how the physical algebra can be implemented efficiently in a distributed system using only a single efficient primitive: range scans over partitioned sorted maps.  We find this primitive to be nearly universally supported across RDBMS, NoSQL systems, linear algebra libraries, file-based systems, and others.

%We envision LARA as a logical successor to MapReduce, a minimal kernel that provides ``just enough'' structure to afford reasoning without sacrificing programmability.

%\begin{enumerate}
%	\item[Adv. 1:] Optimization using properties from LA and RA.
% * <billhowe@cs.washington.edu> 2017-01-13T05:38:04.757Z:
% 
% > Optimization using properties from both LA and RA. 
% Example?
% 
% ^.
%    \item[Adv. 2:] An abstraction for simultaneously implementing LA and RA.
%    \item[Adv. 3:] Different programming models on one system.
%	\end{enumerate}
%   \begin{enumerate}
   
The object of \lara{} is the \emph{Associative Table}, 
a data structure capturing core properties of relations, tensors, and key-values.
The operators of \lara{} are ext (flatmap), join (horizontal concatenation), and union (vertical concatenation).

Our contributions are: 
\vspace{-.1em}
\begin{enumerate}
\itemsep0em 
	\item A minimal logical algebra, \lara{}, to unify LA and RA. % over a single object type.
%     Say something about the proofs?
    %in both reasoning and guiding implementation. \dylan{proof that \lara{} subsumes LA and RA?}\bill{I complained earlier, but I like this point now.  Just a little bit more than MR, and we can directly capture RA and LA, plus a lot of rewrite rules.  With MR, you really don't have any rules out of the box.}
    \item A physical algebra over sorted partitioned maps that exposes low-level optimization opportunities.
    \item A system, \laradb{}, implementing the \lara{} abstractions on the Apache Accumulo database.
    \item An evaluation showing that on representative tasks, \laradb{} is competitive with a hand-coded MR implementation at scale, and far faster at small scales. 
    \item An evaluation of \laradb{} on a more complex sensor validation task that demonstrates the kinds of optimizations exposed by the \lara{} abstractions.
\end{enumerate}\vspace{-.7em}%

\section{Related Work}

MapReduce promoted a minimalist approach to distributed programming, but had no real capabilities for reasoning over and optimizing the resulting programs.  As a result, a spate of SQL-on-Hadoop projects emerged in the first few years.  Since that early period, a number of projects have proposed more refined approaches 
that balance flexibility for the programmer and optimizability by the system.  
% (e.g. HadoopDB \cite{a}, Tenzing \cite{Chattopadhyay2011TenzingAS}) 

Fegaras et al. superimpose three operations atop MapReduce: cmap (abbreviation of concat-map, which implements flatmap), groupBy, and join \cite{fegaras2012optimization}.  These operators expose optimization opportunities, but the authors do not explore the relationship between these operations and linear algebra.  

Elgama et al proposed a variety of sum-product optimizations and operator fusion techniques for SystemML called SPOOF by representing matrix equivalences in RA and applying relational optimizations~\cite{elgamal2017spoof}.  This approach is similar to our own but focuses on developing a single, tightly coupled system involving heavy use of code generation, as opposed to our goal of providing a general abstraction that can be naturally implemented in many contexts.

Kunft et al described a vision for optimization involving both linear and relational algebra equivalences, but did not describe an implementation of the ideas~\cite{kunft2016bridging}.  Our paper was inspired in part by this work; we use an adaptation of their running example in our experiments.

Palkar et al proposed Weld, a common runtime that replaces the runtime of libraries like Spark, Pandas, and NumPy in order to optimize within and across them \cite{palkar2017weld}.
Weld's algebraic basis is an adaptation of map and reduce termed ``loops and builders''.
We designed \lara{} with one step more structure
by differentiating ``horizontal'' from ``vertical'' group-by,
in order to obtain greater reasoning capability.

% emphasizes reducing inefficiencies between operations rather than within operations, relative to hand-optimized programs~\cite{palkar2017weld}.  Our emphasis is to identify and implement the abstractions that enable expressing and reasoning about these workflows rather than solely improving performance using existing libraries and systems.

% Fabregat et al propose a domain-specific compiler for linear algebra operations, but do not emphasize reasoning across a unified relational and linear algebra \cite{fabregat2012domain}. 

% the Dxter optimizer - tensor contractions over MPI Collectives
Marker et al built a cost-based optimizer, DxTer,
and applied it to the task of tensor contractions in MPI environments
\cite{marker2015dxter}.
Though DxTer lays a foundation for RA-style reasoning, 
they focused solely on the domain of dense LA.

%the other RA alternative algebra papers --- van Emden 

Crotty et al developed Tupleware, a cluster programming environment emphasizing code generation and stateful analytics, but the emphasis is on low-level programming idioms rather than marrying logical and physical abstractions~\cite{crotty2015tupleware}.
%or An Architecture for Compiling UDF-centric Workflows \cite{crotty2015architecture}

Rheinlander et al proposed a logical optimizer for UDF-centric dataflows called SOFA~\cite{rheinlander2015sofa}.  SOFA emphasizes properties of UDFs to facilitate optimizations.  Our approach is complementary, but we focus on properties that allow reasoning about LA, specifically semiring structures.
% SOFA emaphasizes both programmer-supplied and automatically-derived UDF properties

%\url{http://www.cockroachlabs.com/blog/sql-in-cockroachdb-mapping-table-data-to-key-value-storage/}
%Implements a relational table inside a key-value store. The primary index only goes from the primary key. Has other indexes.
% \item SQL++ --- nested relational algebra captured lots of SQL and NoSQL systems

Semirings are the main focus of Associative Arrays,
a data structure generalizing LA's sparse matrices 
whose operations were shown to subsume RA \cite{kepner2016assocdb}.
\lara{} and associative arrays share design choices such as sparsity and pluggable $\oplus$ and $\otimes$.
% but ultimately \lara{} takes a key-value approach.
Ultimately, \lara{} compromises between associative arrays (heavily LA-based) and relations (heavily RA-based).

% Kepner et al aim to unify RA and LA with associative arrays,
% a data structure generalizing sparse matrices over semirings with operations strongly resembling LA \cite{kepner2016assocdb}.
% Our \lara{} algebra took inspiration from associative arrays,
% but ultimately takes a more minimal key-value approach.
% % that affords simpler optimization and implementation.
% % with named dimensions in the form of associative tables.

% when all attributes are restricted to Boolean type, all attributes are keys
Lattices were also posed as a basis for RA
via two operators: natural join and inner union \cite{spight2006first}.
% Spight et al reduced RA to two operators, natural join and inner union,
% and showed how they fit a lattice structure \cite{spight2006first}.
The relational lattice is a special case of \lara{},
especially w.r.t. Section \ref{sProps}'s distributive laws,
when \lara{} is relaxed to allow tables with infinite support %(!)
and restricted to only use key attributes.
% The distributive laws they prove resemble the ones we present in Section \ref{sProps}.
% Although elegant, these relaxations are difficult to implement

\section{Logical Algebra}

% \subsection{Intuition}

% primer on RA relations and LA matrices?
% referring to RA tables as relations, to distinguish from associative tables

% In this section we offer intuition on the design of \lara{}, 
% leading into a description of \lara{}'s operators as well as 

% The \lara{} algebra aims to capture the key-value access pattern of relations (keys) and matrices (row and column indices) while affording the properties useful for quantitative analysis.

%A dataset's keys control how we sort, group, or look up its contents.
%In RA, keys are identified by functional dependencies \cite{codd1972further}, often simply a relation's primary key constraint.
%Any number of attributes form the key (in the case of compound keys) and value.
%In LA, keys take the form of indices that map to matrix values.
%A matrix has any number of indices 
%(0 indices per scalar, 1 per vector, 2 per matrix, 3+ per tensor)
%and can store any number of attributes in its values by packing them together
%(e.g., via compound matrix values).
% As RA and LA's operators act on relations and matrices, they act on key-value multidimensional data.

Our hypothesis underlying \lara{} is that the objects
of RA and LA---relations, scalars, vectors, matrices, and tensors---can be recast into a rectangular representation of multidimensional key-value data.
RA and LA operate on ``rectangular blocks'' of key-value records
by either applying a function ``record-wise'' (think selection, projection, flatmap, function application),
by combining blocks ``horizontally'' (think Cartesian product, tensor product, relational joins), 
or by combining blocks ``vertically'' (think relational union, aggregation, tensor contraction).
These operations are composable;
matrix multiply, for example, is a horizontal operation followed by a vertical one.
By expressing these kernels directly, 
rather than their particular instantiation in RA or LA,
we aim to reason about RA and LA tasks uniformly.
We call our new representation an \textit{associative table}.

Figure \ref{fpA} %\ref{f1}\subref{f1a} 
shows an example table consisting of environmental sensor data.
It has two keys---time and measurement class---that map to measurement values.
%We call them \textit{key attributes} and \textit{value attributes}.
The table can be thought of as a lookup function: 
given a key, return the value it maps to, or the default value ($\bot$, in this table)
if the key has no mapping. 
Default values allow us to model sparse and dense matrices uniformly, for example, 
in which case we would use numeric 0 as the default value.

%one looks for a row in the bottom section of the table 
%(called the table's \textit{support}) for one matching the query key. 
%We return the value matching the given key, if listed in the support,
%or if not we return the default value $\bot$, shown in Figure \ref{fpA} in brackets.

%The `[$\bot$]' is an interpretation of non-stored values 
%as non-measured values, i.e., values that were not recorded by the sensor.
%The default value `0' is often an easier choice to work with, since it is the 
%natural identity or arithmetic addition and annihilator of multiplication,
%but in this case we must distinguish a measured 0 from a non-recorded value $\bot$
%in future calculations.
%In other circumstances one might choose other default values such as $-\infty$, 
%a useful choice for finding shortest paths through a graph.
%This approach generalizes the concept of \textsc{null}
%to an explicit value and semantics, 
%which can be helpful for verifying algorithm correctness.
% \dylan{specifically, the default value is the identity for }
% We found default values a better approach for reasoning during query operations
% than the approach to identify non-stored values with \textsc{null} and use special semantics.

\begin{figure}
\vspace{-.5em}
\centering
% \newsavebox{\fmbox}
% \begin{lrbox}{\fmbox}
% \begin{minipage}{0.9\linewidth}
% Table $A$: Sensor data.
% The horizontal bar separates attribute names and default values [in brackets]
% from the table's support. The vertical bar separates keys from values;
% keys not listed map to the default values.
% \end{minipage}
% \end{lrbox}
%Table $A$
\begin{tabular}{cc|c}
% & & $[\bot]$ \\
$t$ & $c$ & $v$ $[\bot]$ \\
\hline  
440 & \humidity{}    & 38.6  \\
466 & \temperature{} & 55.2  \\
466 & \humidity{}    & 40.1  \\
492 & \temperature{} & 56.3  \\
492 & \humidity{}    & 35.0  \\
528 & \temperature{} & 56.5  \\
\end{tabular}
\vspace{-.5em}
\caption{%\usebox{\fmbox}}
Table $A$ with (time, class, value) sensor measurements.
The default value $\bot$ identifies non-measured values.
%The horizontal bar separates attribute names and default values [in brackets]
%from the table's support. The vertical bar separates keys from values;
%unlisted keys map to default values.
}
\label{fpA}
\vspace{-1em}
\end{figure}

% fashion, mold, forge?
%Default values make associative tables into \textit{total functions} from keys to values.
%Even when the key-space is infinite 
%(as with Figure \ref{fpA}'s integer timestamps),
%the number of keys that map to non-default values is always finite.
%Finite support is a requirement.
%It enables databases to physically store associative tables, since they only need store the table's support.

%As a logical algebra, \lara{} leaves plenty of room for implementations to store an associative table.
%The order of the key attributes, value attributes, and rows of the support are insignificant;
%permuting rows and columns in Figure \ref{fpA} does not affect the %table's identity as an associative table.
%Implementations are free to define an ordering on rows and columns in order to store and process them.
%They may use compression, sorted maps, hash tables with bloom filters, horizontal and vertical partitioning,
%or any number of additional structures and techniques to store an associative table.
%Section \ref{sPhysical} presents one extension of \lara{} 
%to a physical algebra on partitioned sorted maps;
% extensions to hash-based map, relational, array, and other physical environments
%are possible and encouraged in future work.
% Future work might consider how to extend \lara{} to a physical algebra, 
% that specifies data layout in more detail. 

% \newlength{\tsep}
% \setlength{\tsep}{0.5em}
\newcommand{\tsep}[0]{\;}
\newcommand{\dl}{[}
\newcommand{\dr}{]}
\newcommand{\cellshade}{\cellcolor{blue!15}}
\newcommand{\rowshade}{\rowcolor{blue!15}}

\begin{figure*}[t]
\vspace{-.5em}
\centering%
\begin{tabular}{@{}p{.36\linewidth}p{.38\linewidth}@{}} %@{\tsep}|@{\!}p{.22\linewidth}@{}}
\multicolumn{1}{c}{Pseudocode} & \multicolumn{1}{c}{\lara{} Logical Plan} \\ %& \multicolumn{1}{c}{$f, \oplus, \otimes$ Functions} \\
\hline
{\scriptsize $\phantom{1}$1}
$A, B$ = \bload{} `s1', `s2'
	& $A$ = \bload{} `s1'
%     &
% % \multicolumn{1}{m{.25\linewidth}}{
% \multirow{9}{*}{
% \newlength{\vsepcases}
% \setlength{\vsepcases}{-.2em}
% \begin{varwidth}{\linewidth}
% $\;$ \\[.1em] %vertical centering
% bin($t$) = $t - \operatorname{mod}(t,60)$ \\
% \phantom{bin($t$}$ +\, 60 \lfloor \frac{\operatorname{mod}(t,60)}{60}+.5 \rfloor$ \\
% $(v\neq\bot) = \begin{cases}
% 0,\! &\!\!\!\bif v = \bot \\[\vsepcases]
% 1,\! &\!\!\!\botherwise
% \end{cases}$ \\
% plus$_\bbot{}$($v,v'$) = \\
% \phantom{\quad}$\begin{cases}
% v', 	&\!\!\bif v=\bot \\[\vsepcases]
% v, 		&\!\!\bif v'=\bot \\[\vsepcases]
% v+v',	&\!\!\botherwise
% \end{cases}$ \\
% % plus($v,v'$) = $v + v'$ \\
% divide$_\bbot{}$($v, cnt$) = \\
% \phantom{\quad}$\begin{cases}
% \bot, 	&\bif v=\bot \\[\vsepcases]
% v/cnt, 	&\botherwise
% \end{cases}$ \\
% minus$_\bbot{}$($v,v'$) = \\
% \phantom{\quad}$\begin{cases}
% \bot, 	&\!\!\!\!\bif v{=}\bot \lor v'{=}\bot \\[\vsepcases]
% v-v', 	&\!\!\!\!\botherwise
% \end{cases}$ \\
% times$_\bbot{}$ = minus$_\bbot{}$ \\
% \phantom{\qquad} with $\cdot$ in place of $-$ \\[.2em]
% or($v,v'$)\! = $(v\neq0) {\lor} (v'\neq0)$ \\[.2em]
% % = \\
% % \phantom{\quad}$\begin{cases}
% % \bot, 			&\!\!\!\bif v{=}\bot \lor v'{=}\bot \\
% % v \cdot v', 	&\!\!\!\botherwise
% % \end{cases}$ \\
% divMinusOne$_\bbot{}$($v,v'$)= \\
% \phantom{\quad}if($v{=}\bot$) $\bot$ else $v/(v'{-}1)$
% \end{varwidth}
% % }
% } 
\\
\cellshade
{\scriptsize $\phantom{1}$2} \emph{// RA-style (SQL) bin, filter} 
	&\cellshade $A_1$ = \bmap{} $A$ \bby{} 
    \dl{}$v$: if($460 \leq t \leq 860$) $v$ else  $\bot$\dr{} 
%     \begin{tabular}{@{}c@{\;}|@{\;}c@{}}
% %  & $[\bot]$ & $[0]$ \\
% &$v$ \\
% \hline
% ()&if($460 \leq t \leq 860$) $v$ else  $\bot$
% \end{tabular}
\\ \cellshade
{\scriptsize $\phantom{1}$3} $A'$ = \bselect{} bin($t$) \bas{} $t', c,$ avg($v$) 
	\newline
    \phantom{{\scriptsize 13} $A'$ =}       \bfrom{} $A$ \bwhere{} $460 \leq t \leq 860$ 
    &\cellshade $A_2$ = \bext{} $A_1$ \bby{} 
%     \dl{}$t'$: bin($t) \rightarrow v$: $v, cnt$: $v\neq\bot$\dr{} 
    \begin{tabular}[t]{c|cc}
%  & $[\bot]$ & $[0]$ \\
$t'$ & $v$ & $cnt$ \\
\hline
bin($t$) & $v$ & $v\neq\bot$
\end{tabular} \\
\rowshade
{\scriptsize $\phantom{1}$4} 
\phantom{$A'$ =} \bgroupby{} $t', c$
	& $A_3$ = \bagg{} $A_2$ \bon{} $t',c$ \bby{} [$v$: +, $cnt$: +] \\
\rowshade 
{\scriptsize $\phantom{1}$5}
	&
    $A'$ = \bmap{} $A_3$ \bby{} \dl{}$v$: $v / cnt$\dr{} \\
% \rowshade
{\scriptsize $\phantom{1}$6} $B'$ = \bselect{} bin($t$) \bas{} $t', c,$ avg($v$)  \newline
\phantom{{\scriptsize 13} $B'$ =}       \bfrom{} $B$ \bwhere{} $460 \leq t \leq 860$
\newline \phantom{{\scriptsize 13} $B'$ =} \bgroupby{} $t',c$
	& $B' = \dots$   \emph{// repeat above for second sensor} \\
%  &  \\[-.1em]
 \multicolumn{2}{@{}l@{}}{{\scriptsize $\phantom{15}$} \emph{// LA-style ({\footnotesize MATLAB}) mean, covariance of residuals $A' - B'$,\! viewed as dense matrices:}} \\[.1em]
% \multicolumn{2}{@{}p{.73\linewidth}@{\tsep}|}{\makebox[.7\linewidth][s]{\emph{// LA-style ({\footnotesize MATLAB}) mean, covariance of residuals $A' - B'$,\! viewed as dense $t' {\times} c$ matrices:}}} \\
% \emph{// LA-style ({\footnotesize MATLAB}) mean, covariance
% of residuals $A' {-} B'$ viewed as dense $t' {\times} c$ matrices:}\!\!\!\! \\
% \emph{// \;\;\; $C = \ex \left[ (X - \ex X)^\tr (X - \ex X) \right]$} \\
\cellshade
{\scriptsize $\phantom{1}$7} $X = A' - B'$ \;\;\emph{// residuals; $|t'| {\times} |c|$ matrix} 
	&\cellshade $X$ = \bjoin{} $A', B'$ \bby{} [$v$: $-$] \\
% $N = \operatorname{size}(A' \lor B', t')$ \\
{\scriptsize $\phantom{1}$8} $N$ = size($X$, 1) \quad\emph{// \# unique t's; scalar} 
\newline {\scriptsize \phantom{1}9}
\newline {\scriptsize 10}
	& $X_1$ = \bmap{} $X$ \bby{} \dl{}$v$: $v\neq\bot$\dr{} \newline
	$X_2$ = \bagg{} $X_1$ \bon{} $t'$ \bby{} [$v$: any] \newline %$(v\neq0) {\lor} (v'\neq0)$
	$N$ = \bagg{} $X_2$ \bby{} [$v$: +] \\
% \\
\cellshade
{\scriptsize 11} $M$ = mean($X$, 1) \emph{//$c$ means; $1 {\times} |c|$ vector} \newline
{\scriptsize 12} \bstore{} $M$
\newline {\scriptsize 13} 
	&\cellshade $X_3$ = \bmap{} $X$ \bby{} \dl{}$v$: $v$, $cnt$: $v\neq\bot$\dr{} \newline
	$X_4$ = \bagg{} $X_3$ \bon{} $c$ \bby{} [$v$: +, $cnt$: +] \newline
	$M$ = \bmap{} $X_4$ \bby{} \dl{}$v$: $v / cnt$\dr{} \\
{\scriptsize 14} $U = X {-} \operatorname{repmat}(M, N, 1)$\emph{//subtract mean} 
	& $U$ = \bjoin{} $X, M$ \bby{} [$v$: $-$] \\
\cellshade
{\scriptsize 15} $C = U^\tr U / (N-1)$ \qquad\quad\emph{// $c$ covariances;} \newline
{\scriptsize 16} \bstore{} $C$\qquad\qquad\qquad\quad\;\emph{// $|c| {\times} |c|$ matrix}
\newline {\scriptsize 17}
\newline {\scriptsize 18}
	&\cellshade $U_1$ = \brename{} $U$ \bfrom{} $c$ \bto{} $c'$ \newline
	$U_2$ = \bjoin{} $U_1, U$ \bby{} [$v$: $\times$] \newline %$v \cdot v'$
	$U_3$ = \bagg{} $U_2$ \bon{} $c, c'$ \bby{} [$v$: +] \newline
	$C$ = \bjoin{} $U_3, N$ \bby{} [$v$: $v / (v' - 1)$] 
\end{tabular}

\vspace{-.7em}
\renewcommand{\tsep}{\;\;}
\newcommand{\hsep}{\;\;}
\newlength{\myheight}
\setlength{\myheight}{73pt} %85
\subfloat[Table $A_2$]{
\label{fpA2}
% \begin{lrbox}{\fmbox}
\begin{varwidth}[t][\myheight]{\linewidth}\vspace{0pt}
\begin{tabular}{@{}c@{\tsep}c@{\tsep}c@{\tsep}|@{\tsep}c@{\tsep}c@{}}
   & & & $[\bot]$  & [0]  \\
  $t$ & $c$ & $t'$ & $v$ & $cnt$ \\
  \hline  
  466 & \temperature{} & 460 & 55.2 & 1  \\
  466 & \humidity{}    & 460 & 40.1 & 1  \\
  492 & \temperature{} & 520 & 56.3 & 1  \\
  492 & \humidity{}    & 520 & 35.0 & 1  \\
  528 & \temperature{} & 520 & 56.5 & 1  \\
  \end{tabular}
\end{varwidth}
% \end{lrbox}\fbox{\usebox{\fmbox}}
} \hsep
\subfloat[Table $A'$]{
\label{fpA'}
% \begin{lrbox}{\fmbox}
\begin{varwidth}[t][\myheight]{\linewidth}\vspace{0pt}
\begin{tabular}{@{}c@{\tsep}c@{\tsep}|@{\tsep}c@{}}
   & & $[\bot]$ \\
  $t'$ & $c$ & $v$ \\
  \hline  
  460 & \temperature{} & 55.2  \\
  460 & \humidity{}    & 40.1  \\
  520 & \temperature{} & 56.4  \\
  520 & \humidity{}    & 35.0  \\
  \end{tabular}
\end{varwidth}
% \end{lrbox}\fbox{\usebox{\fmbox}}
} \hsep
\subfloat[Table $X$]{
\label{fpX}
\begin{varwidth}[t][\myheight]{\linewidth}\vspace{0pt}
\begin{tabular}{@{}c@{\tsep}c@{\tsep}|@{\tsep}c@{}}
   & & $[\bot]$ \\
  $t'$ & $c$ & $v$ \\
  \hline  
  460 & \temperature{} & -3.1  \\
  460 & \humidity{}    & 1.6  \\
  520 & \temperature{} & -4.0  \\
  520 & \humidity{}    & -0.8  \\
  \end{tabular}
\end{varwidth}
} \hsep
\subfloat[$N$]{
\label{fpN}
\begin{varwidth}[t][\myheight]{\linewidth}\vspace{0pt}
\begin{tabular}{@{}c@{\;}|@{\;}c@{\;}}
\multicolumn{2}{c}{} \\
&$v$ \\
\hline
()&2
\end{tabular}
\end{varwidth}
} \hsep
\subfloat[Table $M$]{
\label{fpM}
\begin{varwidth}[t][\myheight]{\linewidth}\vspace{0pt}
\begin{tabular}{@{}c@{\tsep}|@{\tsep}c@{}}
   & $[\bot]$ \\
  $c$ & $v$ \\
  \hline  
  \temperature{} & 0.4  \\
  \humidity{}    & -3.5  \\
  \end{tabular}
\end{varwidth}
}\hsep
\subfloat[Table $U$]{
\label{fpU}
\begin{varwidth}[t][\myheight]{\linewidth}\vspace{0pt}
\begin{tabular}{@{}c@{\tsep}c@{\tsep}|@{\tsep}c@{}}
   & & $[\bot]$ \\
  $t'$ & $c$ & $v$ \\
  \hline  
  460 & \temperature{} & 0.4  \\
  460 & \humidity{}    & 1.2  \\
  520 & \temperature{} & -0.4  \\
  520 & \humidity{}    & -1.2  \\
  \end{tabular}
\end{varwidth}
}\hsep
\subfloat[Table $C$]{
\label{fpC}
\begin{varwidth}[t][\myheight]{\linewidth}\vspace{0pt}
\begin{tabular}{@{}c@{\tsep}c@{\tsep}|@{\tsep}c@{}}
   & & $[\bot]$ \\
  $c_1$ & $c_2$ & $v$ \\
  \hline  
  \temperature{} & \temperature{} & $0.32$  \\
  \temperature{} & \humidity{}    & $0.96$  \\
  \humidity{}    & \temperature{} & $0.96$  \\
  \humidity{}    & \humidity{}    & $2.88$  \\
  \end{tabular}
\end{varwidth}
}

\caption{Example to compute the mean $M$ and covariance $C$ of sensor residual differences $X$ after filtering and aligning their measurements $A, B$.
The example is given in pseudocode in the left panel involving RA and LA operations.
The bin function is defined as bin($t$) = $t - \operatorname{mod}(t,60) + 60 \lfloor \frac{\operatorname{mod}(t,60)}{60}+.5 \rfloor$.
The right panel shows a translation into the \lara{} algebra. 
% The right panel lists functions referenced in the center.
Figure \ref{fpA} presents sample data for sensor $A$; sensor $B$ is omitted.
The bottom panel displays results calculated from the sample data.
% Pseudocode (left) compiles to the \lara{} algebra (center),
% with functions referenced defined at the (right).
% The computation is given in high-level psueudocode and its translation into the \lara{} logical algebra.
% Sample data surround the code, showing results at various points after starting from the first sensor's data in Figure \ref{fpA}. Second sensor's data omitted.
}
\label{fl}
\vspace{-1em}
\end{figure*}

\subsection{Running Example: Sensor Quality}
To illustrate the \lara{} algebra, 
consider the example task and its translation into \lara{} in Figure \ref{fl}.
% consider Figure \ref{fpA2}, which shows a pseudocode task (left) and a \lara{} implementation (right). 
We adapted this example from Kunft et al \cite[Listing 1]{kunft2016bridging}, in which the authors motivated an RA-LA hybrid query language like \lara{}.

%We now demonstrate the \lara{} algebra in the context of 
%an environmental sensor manufacturer's quality control process.

In this example, 
a manufacturer seeks to calibrate newly produced sensors to a ``gold standard'' sensor.
Each sensor records temperature, humidity, and other environmental data, albeit at  different rates and times.
The goal is to compute means and covariances of the residual difference between new sensors' and the trusted sensor's measurements. 
%The result indicates the first and second-order variations useful for calibrating the new sensor.

\paragraph{Pseudocode}
%We show a high-level pseudocode algorithm for computing the means and covariances 
%of the sensor measurement residuals at the left of Figure \ref{fl}.
%We decompose the algorithm into two phases:
%an ``RA pre-processing phase'' that computes the residual difference $X$ between sensors,
%and an ``LA analysis phase'' that computes the means $M$ and covariances $C$
%of the measurement classes $c$.
%For reference, the covariance is given by 
%$C = \ex \left[ (X - \ex X)^\tr (X - \ex X) \right]$.

%We present the algorithm in terms of SQL with assignment for the RA phase,
%, similar to the MyriaL syntax,
%and \matlab{}-like syntax for the LA phase.
%The choice of syntax here is insignificant;
%any syntax that represents the task is suitable.
%For example, the reference paper \cite[Listing 1]{kunft2016bridging} used 
%Scala set comprehensions for RA
%and SystemML's DML syntax \cite{ghoting2011systemml} for LA.

In line 1, we load two sensors' measurements into associative tables $A$ and $B$.
%We show sample data for sensor $A$ in Figure \ref{fpA}.
%The data's keys are measurement time $t$ and class $c$,
%and its values $v$ have default value $\bot$ 
%(``\textsc{null}'', for non-measured values).
%In the LA section, we view these tables as matrices
%with row indices $t$ and column indices $c$.
In lines 3 and 6, the two sensor streams are filtered to a region of interest and binned to minute intervals.  This task is naturally expressed as a SQL query over a set of records.
%, and The RA phase filters and bins $A$ and $B$ into $A'$ and $B'$.
%Filtering restricts the analysis to a fixed time period.
%Binning coarsens the two sensor's time measurements into 60-second buckets,
%averaging together values that fall into the same bucket.
%The binning step is necessary because the sensors record measurements
%at different times and rates;
%aligning them on a relatively coarse interval enables their comparison.
% \dylan{An LA equivalent of this step is a matrix subset for filtering
% and left-multiplication by a projection matrix for binning.}
In line 7, $A'$ and $B'$ are interpreted as matrices, such that the residuals can be computed directly using element-wise subtraction.  This task is possible to express in SQL, but it would involve a multi-attribute join condition and a new column alias. 
%We compute the residual difference of the aligned measurements in $X$.
% \dylan{RA: inner join. LA: element-wise multiplication with subtraction as the $\otimes$ function.} % err... this is not element-wise addition

In line 8, we assume a MATLAB-style function \textsf{size} that computes the number of unique time bins. 
We use this function to illustrate a different programming style; a simple SQL count distinct query would also suffice.
%The LA phase first counts $X$'s number of rows $N$,
%the number of unique times $t$ that both devices record.
% \dylan{RA: aggregation. LA: matrix reduction}
In line 11, the mean of each attribute $c$ across all $t$ is computed.  Here, the MATLAB-style syntax is quite useful; it is tedious and error-prone in SQL to aggregate many columns in one query.

% The vector $M$ contains the mean value for each measurement class $c$ across all times.
% \dylan{RA: aggregation. LA: matrix reduction}
Finally, calculating covariance\footnote{Covariance is given by $C = \ex \left[ (X - \ex X)^\tr (X - \ex X) \right]$}
consists of subtracting the mean from each measurement into $U$ (using the MATLAB function \textsf{repmat}, which repeats the row vector $M$ to the same matrix shape as $X$), then computing $C = U^\tr U$ and dividing by the time count $N-1$.\footnote{Dividing by $N-1$ forms an unbiased covariance estimator.%, as opposed to dividing by $N$.
}
% Matrix $U$ is the result of subtracting each mean from its value.
% \dylan{RA: join. LA: element-wise addition}
% Matrix $C$ is the final covariance matrix, formed via the matrix inner product $U^\tr U$
% and dividing by $N-1$. % unbiased estimator
% \dylan{RA: join and aggregate and apply. LA: matrix multiply and scalar multiplication.}

\paragraph{Lara Logical Plan}
The right of Figure \ref{fl} presents a translation of the pseudocode into a logical plan expressed in the \lara{} algebra. 
Each line in the \lara{} plan consists of a single operator.  
% Although \lara{} has just three core operators (described next), we use syntactic sugar to simplify the exposition.

Figure \ref{tLaraSummary} lists the three core operators of \lara{}, defined in Section \ref{sec:lara}. Here we provide intuition using our running example. The operators we use in Figure \ref{fl} are \bext{} (apply a function to each record, possibly adding new keys), \bmap{}  (apply a function to each record without changing keys), \bagg{} (relational group by), \bjoin{} (relational join on keys), and \brename{} (relabel an attribute).   \bmap{} and \brename{} are special cases of the core operator \bext{}.  \bagg{} is a special case of the core operator \bunion{}.  \bjoin{} is itself a core operator. 

Each line generally takes the form \textit{<op> <table> \textsc{by} <expressions>}, where \textit{<expressions>} is akin to a \bselect{} clause in SQL with aggregate and arithmetic expressions.  
In this example functions handle the default value $\bot$ the same as \bnull{}, 
but a crucial concept in \lara{} is that value attributes may have different default values. %,
% such as the default value 0 for attribute $cnt$ in table $A_2$ of Figure \ref{fl}\subref{fpA2}.
% , and each function may handle $\bot$ in a specialized way.  
Table $A_2$ in Figure \ref{fl}\subref{fpA2} has a 0 default value, for example.
Line 3 uses a tableau notation for the table-valued output of its user-defined function;
Equation \ref{eExtEx} shows an example of this function in action.
%All of the functions distinguish $\bot$ (unmeasured values) from 0 (measured 0s), which is important for computing the correct average.
%The plus$_\bbot{}$ function listed at the right shows how we modify addition to accommodate the behavior that $\bot$ act as an identity to plus.

%In \lara{} form, the algorithm is one step closer to that of an implementation
%but still high-level enough to represent RA and LA operations uniformly and without considering data layout and other lower-level properties.
We encourage the reader to trace through the algorithm with the example tables at the bottom of Figure \ref{fl}, which stem from the first sensor $A$'s data in Figure \ref{fpA}
and a second sensor $B$'s data which is not shown.

\subsection{Lara Defined}
\label{sec:lara}
In this section we define associative tables and the three core \lara{} operators.
% , defined via a COBOL-style syntax.
%We first present them with a COBOL-style syntax conducive to writing scripts,
%and then include an alternative algebraic syntax conducive to writing proofs.
Figure \ref{tLaraSummary} summarizes the \lara{} operators and their effect on associative table structure.

% \begin{table*}
% \begin{tabular}{r|ccc}
%  & new $\tup{k}$ & new $\tup{v}$ & new $\supp$ \\
% \hline
% \bunion{} $A, B$ \bby{} $\oplus$ & $\tup{k}_A \cap \tup{k}_B$ & $\tup{v}_A \cup \tup{v}_B$ & $\subseteq \supp(A) \cup \supp(B)$ \\
% \bjoin{} $A, B$ \bby{} $\otimes$ & $\tup{k}_A \cup \tup{k}_B$ & $\tup{v}_A \cap \tup{v}_B$ & $*$ \\
% \bmap{} $A$ \bby{} $f$ & $\tup{k}_A$ & set by $f$ & $\subseteq \supp(A)$ \\
% \bext{} $A$ \bby{} $f$ & $\tup{k}_A$ extended by $f$ & set by $f$ & $*$ 
% \end{tabular}
% \end{table*}
\begin{figure}[t]
\centering
\begin{tabular}{r|cc}
\lara{} Operator & output $\tup{k}$ & output $\tup{v}$ \\
\hline
\bunion{} $A, B$ \bby{} $\oplus$ & $\tup{k}_A \cap \tup{k}_B$ & $\tup{v}_A \cup \tup{v}_B$ \\
\bjoin{} $A, B$ \bby{} $\otimes$ & $\tup{k}_A \cup \tup{k}_B$ & $\tup{v}_A \cap \tup{v}_B$ \\
\bext{} $A$ \bby{} $f$ & $\tup{k}_A$ extended by $f$ & set by $f$ \\
% \bmap{} $A$ \bby{} $f$ & $\tup{k}_A$ & set by $f$ 
\end{tabular}
\caption{Summary of \lara{}'s operators and their effect on table schema,
i.e., the names of key and value attributes.
% Union and intersection refer here to $A$ and $B$'s attribute names; 
For example, union's key names are the intersection of $A$ and $B$'s key names.
Union and join are dual in this respect.
% \bmap{} is a special case of \bext{}.
}
\label{tLaraSummary}
\vspace{-1em}
\end{figure}

\paragraph{Tuples and Associative Tables}
The notation $\tup{a} = [t\colon 135, c\colon \text{temp}, v\colon 55.2]$
defines $\tup{a}$ as a tuple of three elements named $t$, $c$, and $v$.
All tuples have names associated with their values.
Writing tuples $\tup{a}, \tup{b}$ side-by-side denotes their concatenation.
%when $\tup{a}$ and $\tup{b}$ have no names in common.
We use $\pi_X$ to denote tuple projection onto names $X$; for example, 
$\pi_{t}\, \tup{a} = [t\colon 135]$.

%\paragraph{Associative tables}
An associative table $A : \tup{k} \rightarrow \tup{v} : \tup{0}$ is a total function from key attributes $\tup{k}$ to value attributes $\tup{v}$ with default values $\tup{0}$.  
% Each value attribute is associated with a distinguished default value, often denoted $\bot$.

%We write $A: \tup{k} \to \tup{v}: \tup{0}$ to indicate keys $\tup{k}_A$, values $\tup{v}_A$, and default values $\tup{0}_A \in \tup{v}$.
% We also write these as $\tup{k}_A$, $\tup{v}_A$, and $\tup{0}_A$ to distinguish between different tables.
%The symbols $\tup{k}_A$ and $\tup{v}_A$ may also refer to their set of attribute names by abusing notation.

The \emph{support} of $A$ %, written $\supp(A)$, 
is the set of keys that map to non-default values.
Associative tables always have finite support.
The expression $A(\tup{k})$ retrieves the $\tup{v}$ associated with $\tup{k}$ if $\tup{k}$ is in $A$'s support, 
or default values $\tup{0}$ if $k$ is not in $A$'s support.  
%We will refer to the $i$th attribute of the value with a bracket notation: $A(k)[i] = v_i$.  We will write the projection of $A$ onto attributes $K$ as $A[K]$.  We will write the concatenation of a tuple $x$ and a tuple $y$ as $xy$, and the union of a set of attributes $X$ with a set of attributes $Y$ as $XY$.  We will also refer to the keys and values of $A$ as $keys(A)$ and $values(A)$ respectively.  
%The projection of a tuple $t$ onto attributes $A$ will be written $t_A$. %\not\in \supp(A)$.
%% in which the default values are part of the value-space
% Let $A$ and $B$ have types
% \begin{align*}
% &A : \tup{a}, \tup{c} \to \tup{x}, \tup{z} : \tup{0}^x, \tup{0}^z \\
% &B : \tup{c}, \tup{b} \to \tup{z}, \tup{y} : \tup{0}^z, \tup{0}^y 
% \end{align*}
% such that they have key attributes $\tup{c}$ and value attributes $\tup{z}$ in common.

%\paragraph{User-defined functions}
%The symbols $f$, $\oplus$, and $\otimes$ refer to user-defined functions.
%The former is a function on keys and values, $f(\tup{k}, \tup{v})$,
%and outputs an associative table.
%The latter two take and produce individual values, 
%$v \oplus v'$ and $v \otimes v'$.

%These functions must satisfy certain properties, explained in the following sections,
%in order to be used in associative table operations.
%These properties guarantee maintenance of finite support.
%In short, we require that $f$ produce default values on input %default values,
%that $f$ produce tables with finite support,
%that $\oplus$ have the default value as its identity, 
%and that $\otimes$ have the default value as its annihilator.

\paragraph{Union}
As the ``vertical concatenation'' of tables,
%Let $A : X \rightarrow V$ and $B : Y \rightarrow W$ be associative tables.  Then 
the expression
\begin{center}
\bunion{} $A, B$ \bby{} $\tup{\oplus}$
\end{center}
creates an associative table over the shared key attributes of $A$ and $B$, aggregating values that map to the same key.
%is an associative table $C$ with key attributes $X \cap Y$ and value attributes $V \cup W$.  The support of $C$ is the union of $A[X \cap Y]$ and $B[X \cap Y]$.  For each key $k$ in $C$, the value of each  is the aggregation 

%The values of $C$ are computed as the aggregation of all colliding keys and if $v$ is in 

%The result is an associative table $C$ with support $supp(C) = supp(A) \cup supp(B)$, where each key $k$ in $supp(C)$ is associated with a tuple of values $(A(k)[0] \oplus_0 B(k)[0], A(k)[1] \oplus_1 B(k)[1], \dots, A(k)[n] \oplus_n B(k)[n])$ 

%which aggregates the union of $A$ and $B$'s values onto 
%the intersection of their keys.
% We omit the bar over $\oplus$ when clear from context.

Union takes a tuple of user-defined $\oplus$ functions,
one for each value attribute in $A$ and $B$, to sum colliding values.
Collisions are keys from $A$ and $B$'s support that match on their common key attributes.
The collisions are summed via \emph{structural recursion}
\cite{buneman1995principles},
a strategy to sum values pair-wise until a single value is obtained.
For this paper we assume each $\oplus$ is associative and commutative,
which implies that we can sum values in any order.
In general we can relax this assumption
when $A$ and $B$'s keys have a total order.

We require that each $\oplus$ have $A$ and $B$'s default value as its identity:
$0 \oplus v = v \oplus 0 = v$,
which forces $A$ and $B$ to have the same 0.
%This requirement ensures that the union operation preserves finite support  is independent
This requirement ensures that union has the same result independent of whether default values are stored in $A$ or $B$;
extra default values merely add extra 0s.

Often a \lara{} expression takes the union of a table $A$ with an empty table $E_{\tup{k}}$, 
i.e., a table with key attributes $\tup{k}$ and empty support.
Such a union aggregates $A$ onto the key attributes $\tup{k}$.
For this common case we use the shorthand
\begin{center}
\bagg{} $A$ \bon{} $\tup{k}$ \bby{} $\tupoplus$ %:= \bunion{} $A$, empty($\tup{k}$) \bby{} $\oplus$
\end{center}\vspace{-1em}

\paragraph{Join}
As the ``horizontal concatenation'' of tables, the expression
\begin{center}
\bjoin{} $A, B$ \bby{} $\tup{\otimes}$
\end{center}
% which multiplies pairs of values from $A$ and $B$
% in the Cartesian product of values that map from keys
% which agree on $A$ and $B$'s common key attributes.
creates an associative table over the union of $A$ and $B$'s keys.
Join forms the Cartesian product of $A$ and $B$'s values that match on their keys in common,
and multiplies them.
% Its values are formed by multiplying $A$ and $B$'s values
% Its value attributes are the intersection of $A$ and $B$'s values;
% they are computed by multiplying from the Cartesian product of $A$ and $B$'s tuples that agree their common key attributes. 
% Only values present in both $A$ and $B$ are considered.

A tuple of $\otimes$ functions,
one for each value attribute present in both $A$ and $B$,
multiply their values.
The default value of join is the multiplication of $A$ and $B$'s default values.
% $0 \otimes 0'$.  
% Often $A$ and $B$ have the same default values and so $0 = 0'$, but it is not required.

We require that each $\otimes$ 
have $A$ and $B$'s default values as its annihilators:
$0_A \otimes v = v \otimes 0_B = 0_A \otimes 0_B$.
As with union, this requirement ensures that join is independent of
whether default values are stored in $A$ or $B$;
extra default values merely multiply extra 0s.

\paragraph{Ext}
Also known as ``flatmap'', ext is the extension\footnote{Coined by Buneman et al \cite{buneman1995principles},
  we chose the term ``$\ext$'' over ``$\operatorname{flatmap}$'' to emphasize 
  that ext can extend a table's keys.
  It also indicates monadic bind that is monotonic on key types.}
  % It also indicates the lifting of $f$ from a function on tuples to a function on tables, and it implements a monotonic version of monadic bind.} 
of a function $f$ on tuples to a \lara{} operator on tables written 
\begin{center}
\bext{} $A$ \bby{} $f$
\end{center}
% which extends $A$ with the result of applying $f$ to each tuple in $A$'s support.

The $f$ in ext is a user-defined function that returns a new associative table for every tuple.
The keys in the table returned by $f$ append to $A$'s keys in the result.
The values returned by $f$ replace $A$'s values in the result.
% The new table returned by $f$ can have empty support,
% a single row's support, or larger support, so long as it is finite.

We demonstrate this process with an example from our running sensor example.
Take the ext from Figure \ref{fl} line 3. %constructing table $A_2$ 
The action of this ext's $f$ on the second row of Table $A$ is
%listed in Table $A_1$ (which is Table $A$ in Figure \ref{fpA} after filtering out the first row) 
\begin{align}
\label{eExtEx}
f([t\colon 466, c\colon \text{temp}, v\colon 55.2]) = 
\text{
\begin{tabular}{c|cc}
%  & $[\bot]$ & $[0]$ \\
$t'$ & $v$ & $cnt$ \\
\hline
460 & 55.2 & 1
\end{tabular}}
\end{align}
% The ext applies this $f$ to every tuple in Table $A_1$.
% Table $A_2$ in Figure \ref{fl}\subref{fpA2} shows the ext's final result.

We require that $f$ satisfy two properties to be used in an ext.
The tables produced by $f$ must be valid tables with finite support,
and specifically when passed default values, $f$ must produce a table with empty support.
These requirements guarantee that the result of ext has finite support,
and they offer independence from storing default values in $A$;
default values do not produce support in \bext{}'s result.

A common special case of ext produces no additional key attributes.
% In this case, the tables produced by the ext's $f$ are zero-dimensional tables
% with either zero or one rows in their support.
We call out this behavior with the expression
\begin{center}
\bmap{} $A$ \bby{} $f$
\end{center}
% in order to distinguish it from $f$ functions that truly extend $A$'s key attributes.

To illustrate how \bmap{} relates to \bext{},
we show how the \bmap{} in Figure \ref{fl} line 2 would be written as an \bext{}:
\begin{align*}
\label{eMapExt}
&\bmap{}\; A \;\bby{}\;
    \dl{}v\colon \text{if } (460 \leq t \leq 860)\; v \belse \bot\dr{}
\\
\equiv\quad
&\bext{}\; A \;\bby{}\;
\text{
\begin{tabular}{c|c}
 & $v$  \\
\hline
() & if ($460 \leq t \leq 860$) $v$ else $\bot$
\end{tabular}}
\end{align*}

Another common special case renames keys or values.
Renaming is crucial for the correct application of join and union,
whose semantics depend on the common and distinct names of their input's keys and values.
We write rename as
\begin{center}
\brename{} $A$ \bfrom{} $x$ \bto{} $y$
\end{center}
When $x$ is a value attribute, renaming is straightforward;
a map function $f([x\colon v]) = [y\colon v]$ (holding other value attributes constant) 
performs the renaming.
When $x$ is a key,
the expression is shorthand for an \bext{} that adds a new $y$ key
and an \bagg{} that removes the old $x$ key.
The union does not incur aggregation because collisions cannot occur.

Promoting a value to a key is a simple \bext{},
and demoting a key to a value is an \bagg{} that may incur aggregation.

\paragraph{Formal Definitions}
We now present a more concise algebraic syntax for the \lara{} operators 
that is useful for writing identities and proving theorems.
We encourage the reader to use the COBOL-style syntax when writing scripts.

Suppose we have associative tables $A$ and $B$ with types
\[ A : a, c \to x, z : 0^x, 0^z \qquad
B : c, b \to z, y : 0^z, 0^y \]
i.e., where $c$ and $z$ are keys and values common to $A$ and $B$,
and $a, b, x,$ and $y$ are keys and values unique to $A$ or $B$.
Though we write these as individual attributes, 
the definitions hold when these are tuples (e.g. $\tup{a},\tup{c},\tup{b},\tup{x},\tup{z},\tup{y},\tup{0}^x\!\!,\, \dots$).
% Default values of each value attribute are identified by superscript.

In the case of $\otimes$ and join, the common value attribute $z$ 
is allowed to differ in type and default value between $A$ and $B$.
We omit this case to maintain clarity.

Suppose we have the user-defined functions
\begin{align*}
\oplus &: i \times i \to i \;\;\,\text{ for } i \in \{x, y, z\} \\
\otimes &: z \times z \to z' \\
f &: a, c \times x, z \to (k' \to v' : 0')
\end{align*}
In the case of $\oplus$, the definition holds when there is a different $\oplus$ 
for each attribute $x, y, z$.
However, we only write the case when all the $\oplus$ are the same for clarity.

We require that $\oplus, \otimes,$ and $f$ obey the equations
\begin{align*}
\forall i,\;& i \oplus 0^i = 0^i \oplus i = 0^i \text{ for } i \in \{x, y, z\} \\
\forall z,\;& 0^z \otimes z = z \otimes 0^z = 0^z \otimes 0^z  \\
\forall a,c,k',\;& f(a,c, 0^x,0^z)(k') = 0' \\
\forall a,c,x,z,\;& f(a,c,x,z) \text{ has finite support}
\end{align*}
We also have a technical consistency requirement that $f$ produce tables of the same schema (attribute names) $\forall a,c,x,z$.

Given the above, we define the \lara{} operators as %' type and meaning as 
\begin{align*}
&A \joino{} B : a, c, b \to z : 0^z \otimes 0^{z} \\
(&A \joino{} B)(a,c,b) := [z\colon \pi_z A(a,c) \otimes \pi_z B(c,b)] \\
&A \uniono{} B : c \to x,z,y :  0^x, 0^z, 0^y \\
(&A \uniono{} B)(c) := 
	\big[x\colon \bigoplus_a \pi_x A(a,c), \\
 &\qquad   z\colon \bigoplus_a \pi_z A(a,c) \oplus \bigoplus_b \pi_z B(c,y),\,
    y\colon \bigoplus_b \pi_x B(c,y) \big] \\
&\ext_f A : a,c,k' \to v' : 0' \\
(&\ext_f A)(a,c,k') := f(a,c, A(a,c))(k')
\end{align*}
The ``big $\bigoplus_a$'' denotes summation over all values of key attribute $a$;
the sum is always finite since we sum over associative tables which have finite support.
The ext definition can be seen as un-currying the function given by $f(a,c,A(a,c))$.

We use one shorthand notation.
% When $k' = ()$ in the $f$ of an $\ext_f$, we use the notation $\map_f$,
% as in the previous \bmap{} syntax.
When taking a union with an empty table (one with no support),
as in the previous \bagg{} syntax,
we adopt a unary version of union written $\uniono[x] A$,
where $x$ are the key attributes of the empty table.
We also use $\map_f$ for cases of $\ext_f$ that add no new keys.
%rename

\subsection{Properties and Rewrites}
\label{sProps}

% We now discuss properties and rewrite rules of the \lara{} operators.

% \vspace{-.5em}
\paragraph{Lifted Properties}
Some properties from the user-defined functions $\oplus$ and $\otimes$
automatically apply to union and join.
If $\oplus$ or $\otimes$ are associative, commutative, or idempotent, 
then so are $\uniono$ or $\joino$ respectfully.
These follow directly from the definitions.

\paragraph{Distributive Laws}

First we examine the conditions for distributing join over union.
If $\otimes$ distributes over $\oplus$ such that 
$a \otimes (b \oplus c) = (a \otimes b) \oplus (a \otimes c)$, 
then the same law applies to distribute $\joino$ over $\uniono$ such that
$A \joino\, (B \uniono C) = (A \joino B) \uniono\, (A \joino C)$
under two conditions:
$A$ and $B$ must have no keys in common not also present in $C$, 
and $A$ and $C$ must have no keys in common not also present in $B$.
Put more simply, the distributive law requires
$(k_B \Delta k_C) \cap k_A = \emptyset$, where $\Delta$ is symmetric difference.

Next we examine how to push union through join.
The following result follows from the Generalized Distributive Law \cite{aji2000generalized}.
Assuming that $\otimes$ distributes over $\oplus$,
\begin{align*}
(A \joino\, &B) \uniono C = \\
&((\uniono[k_B \cup k_C] A) \joino\, (\uniono[k_A \cup k_C] B)) \uniono\, (\uniono[k_A \cup k_B] C)
\end{align*}

We refer the reader to a previous technical report for proof of the above two laws
\cite{hutchison2016lara}.

\paragraph{Matrix Equations}
To illustrate the ability to reason about non-trivial equivalences in \lara{}, consider the rotation invariance of matrix multiplication inside a matrix trace:
$\trace(ABC) = \trace(BCA)$.
The trace of a matrix, $\trace(A)$, sums its diagonal entries.
We prove the equation's \lara{} analogue on tables
\[ A: i, j \to v : 0 \qquad B: j,k \to v : 0 \qquad C: k,l \to v : 0
\]
To reduce notation, we use subscript $A_{ij}$ in place of $A(i,j)$.
% all with default value 0.
\begin{align*}
&\trace(ABC) \\
=\; &\unionp \ext_{i=l} (A_{ij}B_{jk}C_{kl}) &\text{$\trace$ defn.}\\
=\; &\unionp \ext_{i=l} \unionp[i,l] (\unionp[i,k]\, (A_{ij} \joinp B_{jk}) \joinp C_{kl}) 
	&\text{$ABC$ defn.} \\
=\; &\unionp \unionp[i,l] \ext_{i=l} (\unionp[i,k]\, (A_{ij} \joinp B_{jk}) \joinp C_{kl}) 
	&\text{push $\ext$ into $\union$} \\
=\; &\unionp \ext_{i=l} (\unionp[i,k]\, (A_{ij} \joinp B_{jk}) \joinp C_{kl}) 
	&\text{combine $\union$} \\
=\; &\unionp\, (\unionp[i,k]\, (A_{ij} \joinp B_{jk}) \joinp C_{ki}) 
	&\text{apply ext} \\
=\; &\unionp \unionp[i,k]\, (A_{ij} \joinp B_{jk} \joinp C_{ki}) 
	&\text{distr. $\join\,$ into $\union$} \\
=\; &\unionp\, (A_{ij} \joinp B_{jk} \joinp C_{ki}) 
	&\text{combine $\union$} \\
=\; &\unionp\, (B_{jk} \joinp C_{ki} \joinp A_{ij}) 
	&\text{commute $\joinp$} \\
=\; &\dots \textit{ // reversing the above steps} \\
=\; &\trace(BCA) 
\end{align*}
% \begin{align*}
% &\trace(ABC) \\
% =\; &\unionp \map_{i=l} (ABC) &\text{$\trace$ defn.}\\
% =\; &\unionp \map_{i=l} \unionp[i,l] (\unionp[i,k]\, (A_{ij} \joinp B_{jk}) \joinp C_{kl}) 
%   &\text{$ABC$ defn.} \\
% =\; &\unionp \unionp[i,l] \map_{i=l} (\unionp[i,k]\, (A_{ij} \joinp B_{jk}) \joinp C_{kl}) 
%   &\text{push $\map$ into $\union$} \\
% =\; &\unionp \map_{i=l} (\unionp[i,k]\, (A_{ij} \joinp B_{jk}) \joinp C_{kl}) 
%   &\text{combine $\union$} \\
% =\; &\unionp\, (\unionp[i,k]\, (A_{ij} \joinp B_{jk}) \joinp C_{ki}) 
%   &\text{apply map} \\
% =\; &\unionp \unionp[i,k]\, (A_{ij} \joinp B_{jk} \joinp C_{ki}) 
%   &\text{distr. $\join$ into $\union$} \\
% =\; &\unionp\, (A_{ij} \joinp B_{jk} \joinp C_{ki}) 
%   &\text{combine $\union$} \\
% =\; &\unionp\, (B_{jk} \joinp C_{ki} \joinp A_{ij}) 
%   &\text{commute $\joinp$} \\
% =\; &\dots \textit{ // reversing the above steps} \\
% =\; &\trace(BCA) 
% \end{align*}

%There are many more matrix identities we could write and prove in the \lara{} algebra.
Due to space considerations, we do not include additional proofs of this form, but we have also sketched proofs of all of the simple rules considered in the context of SystemML \cite{elgamal2017spoof}, including 
$\trace(AB)=\operatorname{sum}(A \otimes B^\tr)$,
$\operatorname{sum}(\lambda \otimes A)=\lambda \otimes \operatorname{sum}(A)$,
$\operatorname{sum}(A + B) = \operatorname{sum}(A) + \operatorname{sum}(B)$, and others.
%We invite the reader to prove the \lara{} analog of one of these identities as an exercise.

\subsection{Relationship to RA and LA}
\label{sRelMR}

Figure \ref{tLARALARA} summarizes how each RA and LA operator
can be written as a \lara{} expression.

First we examine RA in Figure \ref{tLARALARA}\subref{tRALARA}.
Selection ($\sigma$) by a predicate $p$ is a map that sends tuples failing $p$ to the default value.
Projecting away value attributes ($\pi$) is also a map;
projecting away keys is treated as an aggregation.
Aggregation ($\gamma$) and relational union ($\union$) are both instances of \lara{} union.
Relational natural join ($\bowtie$) and Cartesian product ($\times$) are \lara{} joins after ensuring that the join attributes are keys that match in name.
General $\theta$-joins can be modeled via $\sigma_\theta(A \times B)$.

Second we examine LA in Figure \ref{tLARALARA}\subref{tLALARA}.
We chose representative operations for LA based on the emerging GraphBLAS standard  \cite{kepner2016mathblas}.
Matrix multiply ($\!\mxm\!$) is a \lara{} join and union,
after ensuring that the correct dimension of the two matrices match in name.
% Element-wise multiply ($\otimes$) is a join, and element-wise addition ($\oplus$) and reduction are unions.
Element-wise multiply ($\otimes$) and addition ($\oplus$) are a join and union.
Matrix reduction is a union.
Matrix sub-referencing by sets of indices ($A(I,J)$)
is a join of $A$ with each set $I$ and $J$,
treating the sets as indicator vectors with value 1 for each present position and default 0 otherwise.
Function application ($f(A)$) is a map.
Transpose ($A^\tr$) is a rename.

The matrix sub-reference translation highlights an interesting property:
joining a matrix to a vector $A \joino v$ \emph{expands} $v$ to the shape of $A$ and multiplies them together. 
Dually, the union $A \uniono v$ \emph{reduces} $A$ to the shape of $v$ and sums them together.
In LA one must manually adjust shapes before these operations.
\lara{} adjusts them automatically.

RA and LA have more advanced operators we do not cover here,
including outer join, difference, division, pivot, masks, and convolution.
These too are expressible in \lara{} \cite{hutchison2016lara}.
% see a previous technical report for details \cite{hutchison2016lara}.

\begin{figure}[t]
\vspace{-6pt}
\centering
\subfloat[RA to \lara{}]{ 
\label{tRALARA}
  \begin{tabular}{c|c}
  RA & \lara{} \\
  \hline  
  $\sigma_p $ & $\map_p $ \\
  $\pi $ & $\map$ or $\union$ \\
  $\times, \join$ & $\join$ \\
  $\gamma, \cup$ & $\uniono$ \\
  \end{tabular}
} \;\;
\subfloat[LA to \lara{}]{ 
\label{tLALARA}
\begin{tabular}{c|c}
  LA & \lara{} \\
  \hline  
  $A \mxm B$ & $\uniono[i,k] (A \joino B)$ \\
  $A \otimes C$ & $A \joino C$ \\
  $A \oplus C$ & $A \uniono{} C$ \\
  reduce($A, \oplus$) & $\uniono[i] A$ \\
  $A(I,J)$ & $A \join I \join J$ \\
  $f(A)$ & $\map_f$ \\
  $A^\tr$ & rename \\ %rename
  \end{tabular}
}
% Removed this section because ext can be more powerful than RA or LA.
% \subfloat[\lara{} to RA and LA]{
%   \begin{tabular}{c|cc}
%   \lara{} & RA & LA \\
%   \hline  
%   $\ext_f A$ & $\sigma$ or $\pi$ or flatmap &  \\
%   $\pi A$ & $\ext A$ or $\union A$ \\
%   $\times, \join$ & $\join$ \\
%   $\gamma, \cup$ & $\uniono$ \\
%   \end{tabular}
% }
\caption{Translation of RA and LA expressions to \lara{},
given tables $A: i,j \to v,\; B: j,k \to v,\; C: i,j \to v$.}
\label{tLARALARA}
\vspace{-.5em}
\end{figure}

\section{Physical Algebra}
\label{sPhysical}

% \bill{can we say something about iterators over sorted maps? It's a model for key-value systems and relational systems.  The operation is fast in practice. parallelizes well.  Binary operations can be implemented as "merge scans."}

In this section we extend \lara{} to a physical algebra atop an abstraction of
\emph{partitioned sorted maps}.
These maps are a model for many implementations, including 
matrix systems (e.g., those that support CSR, CSC, and DCSC \cite{buluc2008representation} storage),
relational systems (e.g., row and column stores),
and NoSQL systems (e.g., BigTable-style \cite{chang2008bigtable} key-value databases).
%Nevertheless, choosing partitioned sorted maps is not compulsory;
%we anticipate implementing \lara{} as a physical level on other data processing models
%such as hash-based maps or other relational and array models in the future.

We call the new physical algebra \plara{}.
We derive it from \lara{} in three steps.
First, we augment the associative table by imposing an order on their key attributes called an \emph{access path}.
Second, we extend the three \lara{} operators with semantics for associative tables with access paths.
Third, we add the operators \bload{} and \bsort{}.
% We also include \bload{} and \bstore{} operators, and we allow the \bload{} operator to specify a subrange of a table to load from.

% We define \plara{} next, and then discuss optimization opportunities and an implementation on the Accumulo database.

% In this section, we define the physical algebra (Section \ref{spLara}), discuss optimization opportunities (Section \ref{spOpt}), and describe an implementation of \plara{} on the Accumulo database (Section \ref{spAccumulo}). 

After defining PLara, we show how PLara admits a number of RA and LA-style optimization opportunities in the context of the sensor example, and we describe an implementation of PLara on the Accumulo database.

%We explain the above three steps in Section \ref{spLara},
%explore a number of optimization opportunities in Section \ref{spOpt},
%and finally discuss an implementation of \plara{} on the Accumulo database in Section \ref{spAccumulo}.
%We evaluate the effect of the optimization opportunities 
%on our sensor example in Section \ref{sExpExpr},
%and we evaluate the competitiveness of our Accumulo implementation in Section \ref{sExpPerf}.

% \begin{figure*}[tb]
% \begin{mathpar}
% \inferrule*[left=T-MJ]{A: [c,a] \\ B: [c,b] }{\bmergejoin{}\; A, B: [c,a,b]}
% % \inferrule*[left=T-MJ]{A: [c,a] \to x,z: 0^x,0^z \\ B: [c,b] \to z,y: 0^z,0^y}{\bmergejoin{}\; A, B: [c,a,b] \to z: 0^z \otimes 0^z}
% \and 
% \inferrule*[Left=T-MU]{A: [c,a] \\ B: [c,b]}{\bmergeunion{}\; A, B: [c]}
% \and
% \inferrule*[Left=T-Ext]{A: [k] \\ f(k): k' \to v' }{\bapplyext{}\, A \;\bby{}\; f: [k,k']}
% \\
% \inferrule*[left=T-Sort]{ }{\bsort{}\; A \;\bto{}\; [k]: [k]}
% \and
% \inferrule*[left=T-ExtOver]{ }{\bapplyext{}\; A \;\bover{}\; [k]: [k]}
% \and
% \inferrule*[left=T-Load]{\text{Table `x' has access path $[k]$}}{\bload{}\; \text{`x'}: [k]}
% \end{mathpar}
% \caption{Access path inference rules. Each rule infers the access path of a \plara{} expression (below the horizontal bar) based on the access paths of its components (above the horizontal bar). The \textsc{t-load} rule loads the schema from a database catalog. Value attributes are omitted since the \lara{} operations already determine them.}
% \label{fpAPInfer}
% \end{figure*}

\subsection{PLara Defined}
\label{spLara}

A \emph{Sorted Associative Table} is an associative table with an order imposed on its key attributes. We refer to the ordering as an \emph{access path}. For example, the access path of Table $A$ in Figure \ref{fpA} is $[t, c]$.  Its type is written $A: [t,c] \to v : \bot$.

%with Access Paths
%At the physical level, we assign a partial order to the key attributes 
%called an \emph{access path}.
%Tuples with the same access path keys exist in the same partition,
%whereas tuples with different access path keys may exist in different partitions.
% Whereas \lara{} associative tables have no order on their key attributes at the logical level, 
% we require such an order at the physical level for the purpose of sorting.
% We call this order an \emph{access path}.

%Access paths provide a refinement of associative table types 
%that place their key attributes in an ordered list.
% If $A$ instead had the access path $[t]$,
% then it would 
% If an operation somehow destroys a sort order, then we allow 
% The sort order may be partial; 
% An alternative access path $[t]$ would indicates that tuples of $A$
% are sorted by time alone, and that all tuples with the same time exist in the same partition.
% We might choose the latter strategy when it is possible to hold in memory 
% all measurement classes for a particular time.

%Given an access path on an associative table $A$,
We model the map's backing store as a row-wise 
% The choice of row-wise layout is not significant
layout of $A$'s tuples sorted by access path 
and partitioned into segments that can be independently processed or stored.
``Split points'' that delineate partitions are chosen by the implementation,
usually in as equal sizes as possible to avoid skew.
%We leave more sophisticated ``2-D'' and higher partitioning schemes to future work.
% However, we require that tuples with the same keys in the access path
% are stored in the same partition.
% This requirement facilitates the implementation of join and union below.
% % Each partition is independent and can be stored and processed on separate machines or cores.

The above scheme performs horizontal partitioning.
Vertical partitioning can be achieved by separate associative tables;
storing $n$ value attributes separately is equivalent to manipulating 
$A_1 \union{} A_2 \union{} \dots \union{} A_n$, where each $A_i$ is a one-attribute table.
More sophisticated 2-D and higher schemes could be designed
but fall outside this paper's scope.

\paragraph{Sort-on-Write}
Writing out an associative table according to an access path sorts and partitions its data as a side effect.  This mechanism is natural for many database implementations, where inserts automatically sort and partition on a clustered index (SQL) or by keys (NoSQL). 
%Thus, re-sorting a table is as simple as re-inserting the table into a map with a new access path.
%Re-sorting in this way is expensive and should be performed as rarely as possible, 
A chief goal of an optimizer is to minimize the number of sort/write operations.
We typeset 
\begin{center}
\bsort{} $A$ \bto{} $[\tup{k}]$
\end{center}
in bold to highlight the performance impact of re-sorting.

% \dylan{talk about fusion?}

\begin{figure*}[tb]
\centering
% Idea: a single row, centered, with spacing, to allow Access Path to flow out
% \subfloat[\lara{} Physical Plan, with access paths 
% and optimization opportunities annotated]{%      %[\myheight]
% \label{fpPhysical}
\begin{varwidth}[t]{1\linewidth}\vspace{0pt}
\begin{tabular}{@{}l@{\;\;}ll@{}}

% \noindent 
% def $\operatorname{bin}(t)=t-\operatorname{mod}(t,60)+60 \left\lfloor \frac{\operatorname{mod}(t,60)}{60}+.5 \right\rfloor$ \\
% def plus($v,v'$) = $v + v'$ \\
% def minus($v,v') = v - v'$; def or($v, v') = v \lor v'$ \\
% \\
\multicolumn{3}{c}{\lara{} Physical Plan $\qquad\qquad\quad$ Access Path $\qquad\qquad\qquad\qquad\quad$ Optimizations $\qquad\qquad$}\\
% \multicolumn{1}{c}{\lara{} Physical Plan} & 
%   \multicolumn{1}{>{\centering}b{\widthof{$[t',c,$}}}{$\!\!$Access \newline Path} & 
%     \multicolumn{1}{c}{Optimizations}\\
\hline
{\scriptsize $\phantom{1}$1} $A$ = \bload{} `s1'  &$[t,c]$ 
  &(E) Encode numeric attributes in packed byte form\\
\rowshade 
{\scriptsize $\phantom{1}$2} $A_1$ = \bapply{} $A$ \bby{} \dl{}$v$: if($460 \leq t \leq 860$) $v$ else $\bot$\dr{} &$[t,c]$ 
  & (F) Push filter into \bload{} `s1' \bfrom{} 460 \bto{} 860\\
\rowshade 
{\scriptsize $\phantom{1}$3} $A_2$ = \bapplyext{} $A_1$ \bby{} \dl{}$t'$: bin($t) \rightarrow v$: $v, cnt$: $v\neq\bot$\dr{} & $[t,c,t']$ &  \\
\rowshade 
{\scriptsize $3.5$} $A_{20}$ = \bsort{} $A_2$ \bto{} $[t',c,t]$ &  $[t',c,t]$ 
 	&\multirow{-2}{*}{\cellshade (M) Eliminate \bsort{} by $t'$: bin($t$) monotone in $t$} \\
\rowshade 
{\scriptsize $\phantom{1}4$} $A_3$ = \bmergeagg{} $A_{20}$ \bon{} $t'\!,c$ \bby{}$\!$ [$v$: +, $cnt$: +] & $[t',c]$ & \\
\rowshade 
{\scriptsize $\phantom{1}5$} $A'$ = \bapply{} $A_3$ \bby{} \dl{}$v$: $v / cnt$\dr{} & $[t',c]$ & \\
% $A_3$ = \bmap{} $A_2$ \bby{} \dl{}$cnt: cnt \neq 0$\dr{} &$[t',c]$\\
% \\ %, B_3
{\scriptsize $\phantom{1}6$} $B' = \dots$   \emph{// repeat above for second sensor} & $[t',c]$\\
% \\
\rowshade 
{\scriptsize $\phantom{1}7$} $X$ = \bmergejoin{} $A', B'$ \bby{} [$v$: $-$] &$[t',c]$
%   &(Z) Discard entries with value 0\\
	& (P) Propagate $A, B$'s partition splits throughout \\
% $O$ = UNI\bon{} $A_3, B_3, \operatorname{empty}(t')$ \bby{} or($cnt$) &$[t']$\\ %idempotent
% $N$ = \bagg{} $O$ \bby{} plus($cnt$) &$[]$ \\
%  & & (P) Propagate $A, B$'s partition splits throughout\\
{\scriptsize $\phantom{1}8$} $X_1$ = \bapply{} $X$ \bby{} \dl{}$v$: $v\neq\bot$\dr{} &$[t',c]$\\
{\scriptsize $\phantom{1}9$} $X_2$ = \bmergeagg{} $X_1$ \bon{} $t'$ \bby{} [$v$: any] &$[t']$\\
{\scriptsize $10$} $N$ = \bagg{} $X_2$ \bby{} [$v$: +] &$[]$
  &(C) Store scalar $N$ at client instead of a table\\
\rowshade 
{\scriptsize $10.5$} $X_0$ = \bsort{} $X$ \bto{} $[c,t']$ & $[c,t']$ 
	&(Z) If $M, C$ relaxed to \emph{sparse} matrix interpretation, \\
% \\
\rowshade 
{\scriptsize $11$} $X_3$ = \bapply{} $X_0$ \bby{} \dl{}$v$: $v, cnt$: $v\neq\bot$\dr{} &$[c,t']$ 
  &\phantom{(Z) }identify $\bot$ with $0$, discarding 0-valued entries  \\
\rowshade 
{\scriptsize $12$} $X_4$ = \bmergeagg{} $X_3$ \bon{} $c$ \bby{} [$v$: +, $cnt$: +] &$[c]$
  &\phantom{(Z) }in $X_3$ and all following tables  \\
\rowshade 
{\scriptsize $13$} $M$ = \bapply{} $X_4$ \bby{} \dl{}$v$: $v/cnt$)\dr{} &$[c]$ 
  &(D) Defer $X_3, X_4, M$ to future scans on $X_0$, \\
\rowshade 
{\scriptsize $13.5$} \bstore{} $M$ &$[c]$ 
	&\phantom{(D) }eliminating write-out of $M$ \\
{\scriptsize $14$} $U$ = \bmergejoin{} $X_0, M$ \bby{} [$v$: $-$] &$[c,t']$
  &(R) Reuse $X_0$ data source (common sub-expression)\\
{\scriptsize $14.5$}  $U_0$ = \bsort{} $U$ \bto{} $[t',c]$ & $[t',c]$
  &\phantom{(R) }($U_2$ has a similar sub-expression below)\\
% \\
% $U_1$ = \bapply{} $U_0$ \bby{} \dl{}$v$: $v\neq0$\dr{} &$[t',c]$\\
% $U_2$ = \bmergeagg{} $U_1$ \bon{} $t'$ \bby{} or($v$) &$[t']$\\
% $N$ = \bapply{} $U_2$ \bby{} plus($v$) &$[]$\\
% \\
\rowshade 
{\scriptsize $15$} $U_1$ = \brename{} $U_0$ \bfrom{} $c$ \bto{} $c'$ &$[t',c']$
  &(S) $U^\tr U$ is symmetric; only compute upper triangle \\
\rowshade 
{\scriptsize $16$} $U_2$ = \bmergejoin{} $U_0, U_1$ \bby{} [$v$: $\times$] &$[t',c,c']$
  &\phantom{(S) }via \bapply{} filter $c \leq c'$ \\
\rowshade 
{\scriptsize $16.5$} $U_{20}$ = \bsort{} $U_2$ \bto{} $[c,c',t']$ & $[c,c',t']$
  &(A) Push sum of partial products into $U_{20}$ compaction and\\
\rowshade 
{\scriptsize $17$} $U_3$ = \bmergeagg{} $U_{20}$ \bon{} $c, c'$ \bby{} [$v$: +] &$[c,c']$
  &\phantom{(A) }flush; assume no repeated writes due to server failure \\
\rowshade 
{\scriptsize $18$} $C$ = \bmergejoin{} $U_3, N$ \bby{} [$v$: $v/(v'-1)$]  &$[c,c']$
  &(D) Defer $U_3, C$ to future scans on $U_{20}$,\\
\rowshade 
{\scriptsize $18.5$} \bstore{} $C$ &$[c,c']$
  &\phantom{(D) }eliminating final pass
\end{tabular}
\end{varwidth}

\caption{\plara{} physical plan for Figure \ref{fl}'s example,
with shading and line numbers matching the logical plan.
Each line is annotated with its access path. %(i.e. key sort order and partitioning) of its result.
Optimization opportunities are listed at the right;
their effect is quantified in Figure \ref{fOptPerf}.}
\label{fp}
\vspace{-.5em}
\end{figure*}

\paragraph{Sorted Join, Union, Ext}
We assume a single primitive for reading data at the physical level: an efficient \emph{range scan} over the keys of a partitioned sorted map.  
% NoSQL, SQL, and LA systems all typically provide this primitive. 
%that invoke \textit{range iterators} 
%over a sorted stream of data from each partition. % in a given access path range.
Range scans are often implemented as \emph{range iterators} that execute user-defined code,
including filters, transforms, and aggregations,
on streams of data.  

% Partitioned sorted maps have two operations:
% unsorted inserts and sorted parallel range scans.
% Range scans involve a ``range iterator'' that processes each partition 
% in a given range as a sorted stream with user-defined code.
% % These iterators execute arbitrary code,
% % which in our case is code that implements the \lara{} operators.

In previous work we have shown how to re-purpose range iterators,
normally designed for single-table parallel scans,
to multi-table computation \cite{hutchison2015graphulo,hutchison2016graphuloalg}.
% We implemented the ability to scan additional tables
% as well as to insert into other tables during a parallel scan.
This approach enables us to implement the \lara{} operators inside range iterators.

Join and union take the form of \textit{merge-scans}:
range scans on one table that themselves scan matching entries from another.
Processing tables in this way is efficient 
when both tables are sorted on the attributes to be merged;
if not, one must re-sort the input tables prior to the merge-scan.
% Re-sorting means re-inserting a table back into the map
% on a new access path. %\footnote{We do not consider in-memory sorts because the maps }

% Specifically we implement join as \textit{merge-join} and union as \textit{merge-group}.
Specifically we implement \bjoin{}, \bunion{}, and \bagg{} as
\begin{center}
\bmergejoin{} $A, B$ \bby{} $\tup{\otimes}$ \\
\bmergeunion{} $A, B$ \bby{} $\tup{\oplus}$ \\
\bmergeagg{} $A$ \bon{} $[\tup{k}]$ \bby{} $\tup{\oplus}$
\end{center}
\bext{} $A$ \bby{} $f$ maintains the same syntax in \plara{}.

The access path of each operation is as follows.
Assume $A$ has access path $[c,a]$, and $B$ has access path $[c,b]$.
% ($c$ are the keys $A$ and $B$ have in common).
\bmergejoin{} has access path $[c,a,b]$.
\bmergeunion{} has access path $[c]$.
\bmergeagg{} has access path $[\tup{k}]$.
If $f$ produces tables sorted on $[k']$,
then \bext{} $A$ \bby{} $f$ has access path $[c,a,k']$.
% Optimization (M) in the next section shows how
% this access path may be re-ordered if $f$ is known to be monotonic in part of $a$ or $c$.

% Figure \ref{fpAPInfer} shows how to infer the access path of each \plara{} operator from its inputs.
% Rules (T-MJ) and (T-MU) require their input tables $A$ and $B$ sorted on their common keys $c$.
% % These operators require their inputs sorted on exactly the join or group-by attributes, i.e.
% % the attributes common to the input tables,
% % which may be none in the case of a Cartesian product or a full aggregation.
% Rule (T-E\textsc{xt}) appends additional keys produced by $f$ to the end of $A$'s access path.
% Rule (T-\textsc{ExtOverride}) becomes relevant in the next section, 
% when optimization (M) recognizes that the output of an \bext{} can have a different access path due to properties identified in $f$.

The behavior of the these operators is as follows.
\bmergejoin{} $A, B$ takes the Cartesian product of tuples that match on their common keys.
For each match, it streams through $B$'s matching tuples while holding $A$'s matching tuples in memory,
and it applies an $\otimes$ function to each pair.
% Splits are restricted to not split in the middle of a join attribute set.
% In Accumulo, this means putting some attributes in the column qualifier.
% idea: handle these border cases at the client
\bmergeunion{} aggregates tuples by an $\oplus$ for each common key.

% For example, suppose $U$ has the access path $[t',c]$ and $U_1$ has the path $[t',c']$,
% as with the join in line 16 from the sensor example.
% The resulting stream from \bmergejoin{} $U, U_1$ has access path $[t,c,c']$.
% If we were to \bmergeunion{} instead, 
% the resulting stream would have access path $[t]$.

The execution of \bmergeunion{} depends on the properties of the $\oplus$ function.
At a minimum, $\oplus$ must have an identity 0 matching 
the default values of its input tables, or else correctness is not guaranteed.
% \footnote{
% Because implementations are free to store or not store default values,
% the $\oplus$ function may or may not run on the default value.
% Even implementations that do not store default values may store a spurious one
% when adding, for example, $-1 + 1$.
% The only safe way to guarantee union's correctness is to ensure that it behaves the same 
% whether a default value is stored or not.
% }
A basic execution strategy folds $\oplus$ across matching tuples in order on a single partition.
% We may execute any $\oplus$ meeting the identity requirement
% by ensuring each group lies on the same partition and folding the $\oplus$ across pairs of entries from the first to the last element of the group in order.
If $\oplus$ is associative,
then $\oplus$ may run across multiple partitions in parallel,
computing local sums before combining them into a global sum during the next \bsort{}
(see optimization (A) in the next section).
If $\oplus$ is idempotent, 
then $\oplus$ can run more than once on the same tuples, 
which is helpful for guaranteeing correctness when recovering from server failure.
If $\oplus$ is commutative,
then $\oplus$ can run out of order. %, should such a situation arise.
% The merge-group reduces each group to a single tuple via $\oplus$ structural recursion.

% \dylan{talk about the structural recursion, fusion. 
% For $\oplus$, need identity 0. Associativity enables partitioned execution. Idempotence necessary for Accumulo write-at-least-once semantics. 
% Commutativity enables out-of-order execution (not used).
% Fusion up to the point a re-sort required.
% Associative $\oplus$ allows partial (local) sums.}

% \bill{Does a non-monotonic f always imply the need for a sort/write?}
\bext{} $A$ \bby{} $f$ applies $f$ to each tuple,
producing a nested table for each tuple which is immediately flattened.
\bmap{} is similar to \bext{}, but never needs to flatten.
% Each entry in the resulting table 
% producing a table of tuples per input.
% % It does not have an access path requirement and need not modify the access path.
% Additional keys $f$ produces output in sorted order
% and are flat-mapped into the stream of data from $A$.
% % then the keys append to the end of the access path.
% If additional keys are monotonic with respect to some  
% existing keys, then the access path of the resulting stream may have the new key
% inserted before those existing ones, as detailed in rule (M) in the next section.

\bload{} initiates a range scan on an existing table, 
possibly restricted to a sub-range.
Its access path is given by a database catalog.
We also include a \bstore{} operator, implemented as a \bsort{} that does not change the access path. %(Recall sorting is performed by inserting.)

% The previous two paragraphs show how knowing properties of the $\oplus$ and $f$ functions influence their execution inside \lara{} union and ext.
% This strategy for propagating properties from UDFs to their execution is not new; 
% UDF-aware optimizers such as Tupleware \cite{crotty2015tupleware} and SOFA \cite{rheinlander2015sofa} take advantage of UDF properties 
% both manually specified and automatically derived via static analysis.
% We view one contribution of \lara{} as a way of organizing of these UDFs 
% around their structural behavior into join, union, and ext,
% which in turn facilitates reasoning about them in the same way 
% that \lara{} improves over MapReduce.

\paragraph{Physical plan for sensor example}
Figure \ref{fp} presents a line-by-line translation of the \lara{} logical plan from Figure \ref{fl}
into a \plara{} physical plan.
The bulk of the translation is tracking the access path induced by each \lara{} operator
and inserting a \bsort{} where access path requirements are unmet.
This occurs for table $A_3$, which aggregates on $t'$ and $c$ but follows an \bext{} with access path $[t,c,t']$;
table $X_4$, which aggregates on $c$ but stems from $X$ with access path $[t',c]$;
table $U_2$, which joins on $t'$ but stems from $U$ with access path $[c,t']$;
and table $U_3$, which aggregates on $c$ and $c'$ but stems from $U_2$ with path $[t',c,c']$.

\subsection{Physical Optimizations}
\label{spOpt}

\begin{figure*}[t]
\begin{mathpar}
\vspace{-.7em}
\inferrule*[left=(A)]{\bmergeagg{}\; (\bsort{}\; A \;\bto{}\; [k]) \;\bon{}\; k' \;\bby{}\; \oplus}
	{\bsortagg{}\; A \;\bto{}\; [k'] \;\bby{}\; \oplus}
\and
\inferrule*[Left=(F)]{\bload{}\; \text{`x'}: [k] \to v: 0 \\ \bmap{}\; (\bload{}\; \text{`x'}) \;\bby{}\; [v\colon \text{if } (a \leq v \leq b)\; v \belse 0])}
	{\bload{}\; \text{`x'} \;\bfrom{}\; a \;\bto{}\; b}
\and
\inferrule*[Left=(M)]{A: k \to v \\ f(k,v): k' \to v' \\ \text{$f$ monotone in $k$} \\ \bsort{}\; (\bext{}\; A \;\bby{}\; f) \;\bto{}\; [k',k]}
	{\bext{}\; A \;\bby{}\; f \;\bover{}\; [k',k]}
\\
\inferrule*[Left=(Z-Sort)]{\bmap{}\; (\bsort{}\; A \;\bto{}\; [k]) \;\bby{}\; [v\colon \text{ntz}(v)]}
	{\bsort{}\; (\bmap{}\; A \;\bby{}\; [v\colon \text{ntz}(v)]) \;\bto{}\; [k] %\\ f' = f \text{ without }
}
\and\quad
\inferrule*[Left=(Z-Map)]{\bmap{}\; (\bmap{}\; A \;\bby{}\; [v\colon f(v)]) \;\bby{}\; [v\colon \text{ntz}(v)]
	\\ f(\bot)=\bot,\; f(0)=0}
	{\bmap{}\; (\bmap{}\; A \;\bby{}\; [v\colon \text{ntz}(v)]) \;\bby{}\; [v\colon f(v)] %\\ f' = f \text{ without }
}
\and
\inferrule*[Left=(Z-Agg)]{\bmap{}\; (\bagg{}\; A \;\bon{}\; [k] \;\bby{}\; [v\colon \oplus]) \;\bby{}\; [v\colon \text{ntz}(v)]
	\\ \bot \oplus v = v \\ v \neq \bot \Rightarrow 0 \oplus v = v}
	{\bagg{}\; (\bmap{}\; A \;\bby{}\; [v\colon \text{ntz}(v)]) \;\bon{}\; [k] \;\bby{}\; [v\colon \oplus]
}
% \inferrule*[Left=(Z-Union)]{\bmap{}\; (\bunion{}\; A, B \;\bby{}\; [v\colon \oplus]) \;\bby{}\; [v\colon \text{ntz}(v)]
%   \\ \bot \oplus v = v \\ v \neq \bot \Rightarrow 0 \oplus v = v}
%   {\bunion{}\; (\bmap{}\; A \;\bby{}\; [v\colon \text{ntz}(v)]),
%   (\bmap{}\; B \;\bby{}\; [v\colon \text{ntz}(v)]) \;\bby{}\; [v\colon \oplus]
% }
\and
\inferrule*[Left=(Z-Join)]{\bmap{}\; (\bjoin{}\; A, B \;\bby{}\; [v\colon \otimes]) \;\bby{}\; [v\colon \text{ntz}(v)]
	\\ \bot \otimes v = \bot \\ v \neq \bot \Rightarrow 0 \otimes v = 0}
	{\bjoin{}\; (\bmap{}\; A \;\bby{}\; [v\colon \text{ntz}(v)]),
    	(\bmap{}\; B \;\bby{}\; [v\colon \text{ntz}(v)])
    \;\bby{}\; [v\colon \otimes]
}
\vspace{-.75em}
\end{mathpar}
\caption{A sample of rewrite rules.
Rule (A) pushes \bmergeagg{}s into \bsort{};
(F) pushes filters into \bload{};
(M) eliminates \bsort{} after a monotonic \bext{};
the (Z-) rules push discarding zeros.
The ``null-to-zero'' function is ntz($v$) = if ($v=\bot$) 0 else $v$.
% The bottom derivation tree applies the (Z) rules to lines 16.5--18 of Figure \ref{fp}.
}
\label{fpOpt}
\vspace{-.9em}
\end{figure*}

% The \plara{} algebra admits a variety of physical optimizations.
Figure \ref{fpOpt} illustrates a few optimization rules on the \plara{} algebra;
Figure \ref{fp} pinpoints where these and other rules apply to our running sensor example.
We evaluate the impact of these optimizations in Section \ref{sExpExpr}.

Some of the most important optimizations act on \bsort{} operations.
Rule (A) pushes aggregations that run after a \bsort{}
into the \bsort{} itself.
We call the fused operation \bsortagg{}.
Speedup from fusing aggregation into sorting can be dramatic,
since the implementation can compute partial sums 
which reduces the burden of sorting and storage.

%In our Accumulo implementation described in the next section,
%the \bsortagg{} operations takes the form of placing the \bagg{}'s $\oplus$ function on the flush and compaction iterators. 
%These iterators run when Accumulo dumps newly inserted data from memory to a Hadoop file,
%decreasing entries written to disk, 
%and when Accumulo merges Hadoop files of data together,
%decreasing entries re-written.
%The optimization assumes that no repeated writes occur,
%which can happen in the event of server failure and recovery,
%since the addition $\oplus$ function is not idempotent.

Rule (M) eliminates \bsort{}s after an \bext{} 
when they are unnecessary. % because the \bext{} function is monotone in certain key attributes.
Normally additional key attributes produced by an \bext{} append to the end of its input's access path. %, assuming the \bext{} function $f$ outputs in sorted order.
Moving new attributes up in the access path past existing key attributes requires re-sorting.
If $f$ is \emph{monotone} with respect to existing key attributes, however, the new key attributes may be promoted past those existing ones without sorting.
Here, monotone means that $k_1 \leq k_2 \Rightarrow f(k_1) \leq f(k_2)$,
using $f(k)$ to refer to the keys of the tables produced by $f$.

Rule (F) pushes filter operations into the \bload{} statements that start a range scan.
These filter operations restrict data to a range of keys on a prefix of the loaded table's access path.
The result restricts scanned data to the desired range, 
as opposed to naively scanning all  data and post-filtering.

Rules (Z-\textsc{Sort}), (Z-\textsc{Map}), (Z-\textsc{Agg}), and (Z-\textsc{Join}) push ``discarding zeros'' through a \lara{} expression.
These rules generalize to \bext{} and \bunion{}; in fact, they apply at the logical \lara{} level,
but we list them here since they are usually associated with physical storage.
The null-to-zero function---ntz($v$) = if ($v=\bot$) 0 else $v$---changes $v$'s default value from $\bot$ to 0.
The change allows implementations to discard zeros without fear of incorrectness.

In order to apply the (Z) rewrites, 
the inference rules check that the function to push ntz through treats $\bot$ and 0 ``the same''.
For example, read the first as ``if we \bext{} $A$ by $f$ and afterward discard zeros, 
and it holds that $f(\bot)=\bot$ and $f(0)=0$, 
then we can safely discard zeros before the \bext{}''.
% The derivation tree in Figure \ref{fpOpt} applies the (Z) rules to the last three expressions of the sensor example in Figure \ref{fp}, 
% showing how these rules can be used in practice.
% The \underline{\bmap{}} that discards zeros is underlined for emphasis.
% If we were to continue the derivation,
% a natural next step is to eliminate the ntz over $N$ since $N$ is never $\bot$.

We now discuss a few rules not listed in Figure \ref{fpOpt}.
Rule (S) leverages symmetry of the inner product computed in lines 15--17:
$(U^\tr U)^\tr = U^\tr U$. %, which is a \brename{} identity in \lara{}.
% Equation \ref{eSymmetric} captures symmetricity in terms of a \lara{} expression.
\lara{} expresses this identity as a \brename{}.
If $C$ is relaxed to restrict its output to its upper triangle,
then (S) could push the filter 
``\bmap{} $C$ \bby{} $c \leq c'$''
up through the plan to the point immediately after line 16
by means of rules in the same style as the (Z) rules.

Rule (D) defers the last pass before a \bstore{}
to scans on the last materialized table, i.e., the last \bsort{} result.
This rule %does not affect the plan itself but merely 
partitions the plan into parts computed ``eagerly'' 
and parts computed ``lazily''.
The performance impact to future scans of deferring the last computation
is usually minimal, since \bsort{}s are never deferred
and so the deferred operations can be streamed.
In Figure \ref{fp}, lines 11--13 defer to scans on $X_0$,
and lines 17--18 defer to scans on $U_{20}$.

Rule (E) encodes numbers in a packed byte format.
Like (Z) and (S), (E) involves a change in the output that,
if allowed, can be pushed through the computation, 
in this case all the way to the original data sources $A$ and $B$.

Rule (R) is a form of common sub-expression elimination.
In the Accumulo implementation,
(R) entails re-using a single range iterator to serve two separate data streams.

% Rule (C) stores scalars at a client rather than a table.

Rule (P) acts on the table splits that partition data.
It pre-splits new tables using the splits of existing tables.
Pre-splitting tables improves insert performance by increasing parallelism 
before the implementation splits data on its own.

Additional optimizations are possible.
For instance, we might forgo sorting in favor of hash-shuffling
when correct to do so, just as Tenzing employed %over MapReduce
\cite{Chattopadhyay2011TenzingAS}.

\subsection{Accumulo Implementation of LaraDB}
\label{spAccumulo}

We implemented \plara{} on the architecture of Google's BigTable \cite{chang2008bigtable},
a design that closely resembles \plara{}'s sorted partitioned map abstraction.
Operations in the BigTable architecture consist of inserts and range scans.
During scans, the user can execute arbitrary code in the form of \textit{iterators}
that run server-side as data streams from each partition in parallel.
%These range iterators are general enough to implement \lara{}'s operators.
Iterator code can even initiate scans on or write entries to additional tables,
a fact we previously exploited in the Graphulo matrix math library  \cite{hutchison2015graphulo,hutchison2016graphuloalg}.

BigTable's range iterators suffice to implement \lara{}.
In particular, we implemented \plara{} on Apache Accumulo, 
an open-source adaptation of BigTable's design.
However, we emphasize that our implementation applies just as much to other BigTable systems, 
including Hypertable and Apache HBase,
and that we see no fundamental barriers to implementing \lara{} atop other systems
with some concept of key and value, including relational and matrix systems.
Even nested relational systems for JSON-like data fit into \lara{},
either by flattening or new indexing techniques \cite{Shukla2015SchemaAgnosticIW}.

%%%%%% Dylan: I think I should omit details of how the data is exactly stored as key-value entries. As much as I want to include all the gory details, it has little value to the reader.
% We store an associative table as a collection of Accumulo key-value entries.
% Each Accumulo entry stores one value attribute.
% The Accumulo entry's key is composed of the key attributes and the name of the value attribute;
% the Accumulo entry's value is simply the value attribute.
% % Each entry contains the key attributes and the name of the value in the Accumulo row and a single value in the Accumulo value.
% Thus, one ``row in the associative table'' consists of an entry for each value attribute in the row.
% Encoding associative tables in this way is more efficient than one might think,
% since Accumulo uses relative-key compression,
% and it allows for fine-grained inserts and scans.

For this prototype implementation, 
we chose a simple model that stores the first key, subsequent keys, and values in the Accumulo row, column qualifier, and value, respectively.
Keys are stored (and sorted) according to the table's access path.
% We leave the challenge of exploiting other Accumulo components---column family, visibility, or timestamp---to future \plara{} extensions.
We coded 
\bext{}, \bmergejoin{}, and \bmergeunion{}
as iterator fragments linked by the Graphulo library.

\section{Experiments}

\subsection{Sensor Optimization Experiment}
\label{sExpExpr}

In this section we conduct an experiment with two goals:
to assess \lara{}'s ability to express a complex computation with elements of both RA and LA, 
and to measure the impact of optimizations that \lara{} affords on this computation.
We implemented each optimization manually;
building an optimizer that applies them automatically is future work.

% to measure the optimization opportunities afforded by \plara{}
% on our Accumulo implementation.
% Our goal is to assess \lara{}'s 
% ability to express a complex computation
% with elements of both RA and LA, 
% as well as the potential for expressing useful optimizations.
% We implemented each optimization manually;
% building an optimizer that applies them automatically is future work.

The experimental task is the sensor quality control plan detailed in Figure \ref{fp}.
We obtained 1.5 months' data from two ``Array of Things''
sensors \cite{arrayOfThings} managed by Argonne National Laboratory.
The raw data amounts to 1.2 GB;
however, this reduces to about 60 MB after parsing, projecting, and storing the data in Accumulo's default compressed format.
We partitioned each sensor's data into 3 day segments.
The plan's filter step restricts analysis to a 30 day period.

We experimented on an Amazon EC2 \texttt{m3.large} cluster 
of 4 workers, 3 coordinators, and 1 monitor machine.
Each has 7.5 GB memory, %(allocating 3 GB to YARN and 3 GB to Accumulo), 
2 vCPUs, and a 30 GB SSD drive.
% The small number of vCPUs limits intra-node parallelism;
% we expect greater per-node performance on nodes with more cores and perhaps more disks.

Figure \ref{fpOpt} plots sensor task runtime
with different optimizations enabled.
At the left we plot the baseline, no optimizations, at 1230 seconds.
We then plot each optimization individually
as well as the combined effect of all optimizations.

Most of the runtime is spent calculating the covariance $C$.
This matches our expectations because computing the inner product $U^\tr U$ 
generates a large number of partial products.
For this reason, optimization (A) yields the greatest performance increase,
since it drastically increases the efficiency of summing partial products.
Without (A), all partial products must be materialized before they can be summed.

Optimizations (D) and (S) both affect the $C$ calculation 
and deliver the next best performance improvement.
(S) eliminates half the computation to compute $C$, 
and (D) defers finishing the summation to future scans.

% Optimization (P) increased parallelism in each step,
% somewhat reducing the effect of worker skew.
% We expect that introducing additional skew-handling techniques as in \cite{hua1991handling} would further improve this optimization.

Other optimizations proved effective but had less impact since they applied less to the covariance bottleneck.
The impact of (Z) depends on the number of zero-valued entries materialized during the $U$ and $C$ computation.
(P) increased parallelism in each step, somewhat reducing worker skew.
(F) sped up the first phase by 4x, decreasing its runtime from 87 to 22 seconds.
(E) and (M) had smaller effects.
% (C) did not merit measurement.

We conclude that \lara{} and \plara{} 
are sufficient to express the sensor quality control computation,
as well as several optimizations useful for impacting performance.

\begin{figure}[t]
\vspace{-.1em}
\includegraphics[width=\linewidth,clip,trim={14px 14px 2px 15px}]{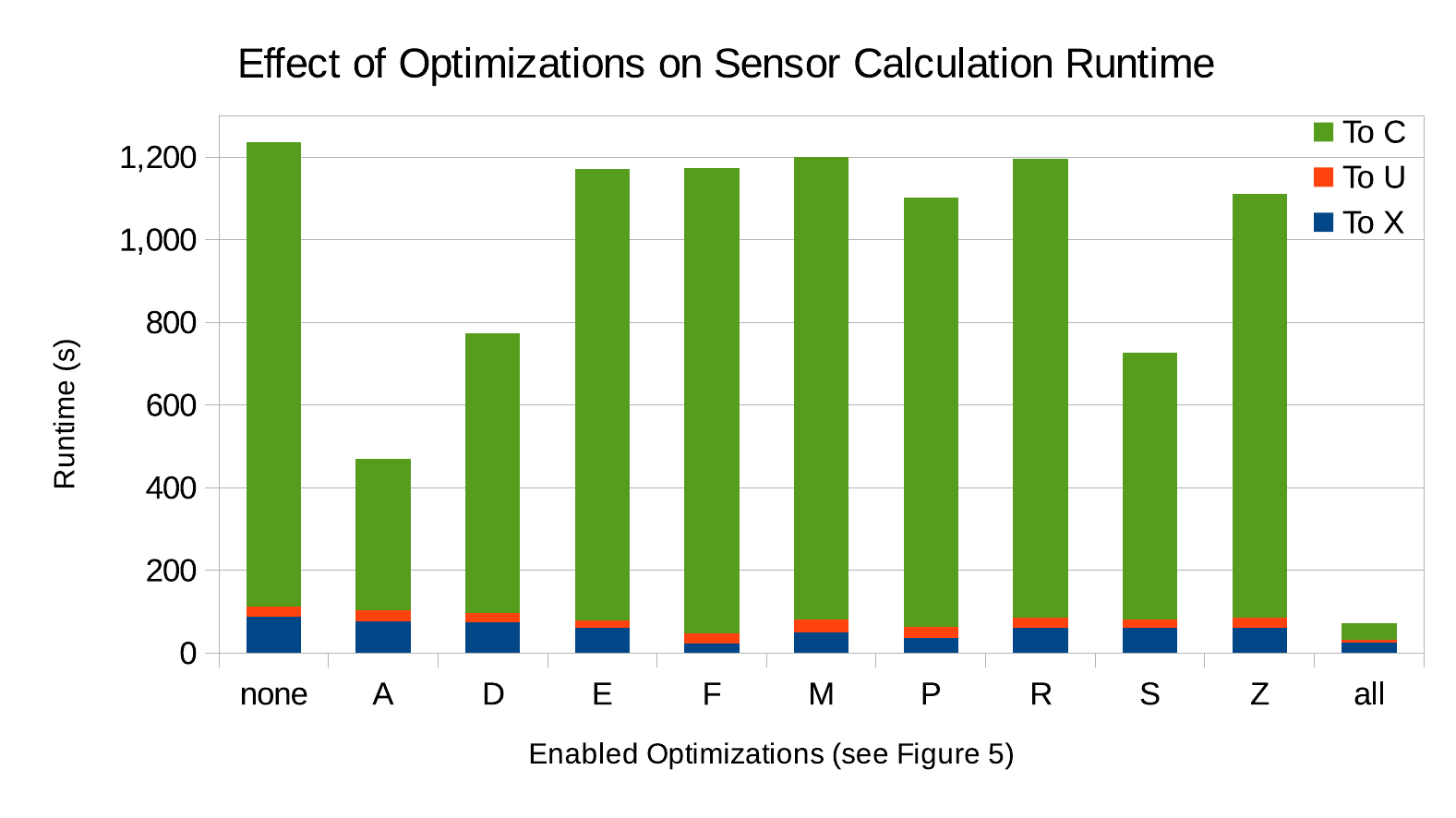}
\caption{Runtime of Figure \ref{fp}'s \plara{} plan
with different optimizations enabled
on one month's data from two sensors.
We decompose runtimes for Figure \ref{fp} into the portion scanning $A$ and $B$ to calculating $X$, from $X$ to $U$, and from $U$ to $C$;
the $C$ calculation dominates runtime.
% We plot each optimization individually against the left, no-optimization baseline.
% The right bar plots runtime with all optimizations enabled.
}
\label{fOptPerf}
\vspace{-1em}
\end{figure}

\subsection{Competitiveness Experiment}
\label{sExpPerf}

In this section we conduct an experiment to test whether \laradb{} 
competes in performance with the analytics engine natively integrated with Accumulo: MapReduce.
% capability of our target platform: the MapReduce/Hadoop engine integrated with Accumulo.

%Specifically we compare \laradb{} to MapReduce.
%MapReduce is well-suited as a comparison system 
%since it is commonly used for analytics on Accumulo data,
%and it has been used for both RA \cite{thusoo2009hive} and LA \cite{seo2010hama,ghoting2011systemml} processing.
%For example, the U.S. government used MapReduce alongside Accumulo 
%to conduct breadth first search on petabyte graphs \cite{burkhardt2015cloud}.
%While other systems such as Spark \cite{zaharia2010spark} arguably supersede MapReduce, 
%our goal is to show \laradb{}'s competitiveness rather than across-the-board top performance.

% In this section we show an experiment comparing matrix multiplication performance
% in \laradb{} with MapReduce. \dylan{explain Graphulo, Accumulo, etc.}
% Our goal is to show that our implementation of \lara{} on Accumulo stands up to existing systems
% used for RA and LA processing.
% We discovered that our implementation not only stands up to MapReduce at scale 
% but also outperforms MapReduce at smaller scales, due to MapReduce's startup overhead.

The task we run is matrix multiplication (MxM).
%on \laradb{} and MapReduce 
%a fundamental building block for both RA and LA.
In terms of RA, MxM consists of a join followed by an
aggregation.
%Join and aggregation are arguably the most complex parts of RA and are often bottlenecks.
In terms of LA, many other LA kernels can be simulated by MxM. For example,
matrix reduction can be realized as multiplication by a vector of 1s,
and matrix subset can be realized as multiplication 
on the left by a diagonal matrix that selects rows 
and on the right by a diagonal matrix that selects columns.
Composition of these kernels lead to more complex graph algorithms
such as triangle enumeration \cite{wolf2015task}, 
vertex similarity, k-truss, and matrix factorization \cite{gadepally2015gabb}.

Because our goal is to compare the performance of the \laradb{} and the MapReduce execution engines,
rather than the difference between two MxM algorithms, 
we wrote the \laradb{} and MapReduce code implementing MxM as similarly as possible.
Both read inputs from and write outputs to Accumulo tables.
Both implement the the MxM $C = A B$ outer product algorithm \cite{hutchison2015graphulo}
on pre-indexed data with $A$ sorted column-major and $B$ sorted row-major.
Both have optimizations (A) and (D) from Section \ref{spOpt} enabled.

The main operational difference between the \laradb{} and MapReduce execution is that
\laradb{} executes inside Accumulo's range scan iterators 
while MapReduce executes as external processes managed by the YARN scheduler.
Specifically, MapReduce performs a reduce-side join \cite{fegaras2012optimization}.

% \dylan{mention that the MapReduce join is known as a reduce-side join [27]}
% but see https://www.semanticscholar.org/paper/Cascading-map-side-joins-over-HBase-for-scalable-Przyjaciel-Zablocki-Sch%C3%A4tzle/b9c652b50b3b3d31f37cc14516fcc49b82552216

% \dylan{Could go into more detail about execution:
% Accumulo code executes the merge-join during a scan and defers the merge-group union to later scans(/flushes/compacts);
% MapReduce mappers align input tables; reducers merge-join; defers merge-group union in the same way.
% Number of reducers did not appear to affect performance significantly.
% Made adjustments to MapReduce execution for correctness.
% I have figures that depict the Graphulo/MapReduce versions, but I don't think this paper is a place for them.}
% We store all components as byte-encoded strings for simplicity.\footnote{Packing the matrix entries as integers would save space and increase performance, but only in a way that benefits MapReduce and Graphulo equally.}

%  I could make the Accumulo code super reliable by writing partial products individually, 
% only allowing deletion (during flush/compact) after all partial products inserted.
% For (j,i,v1)*(j,k,v2), would put the original j after i|k as i|k|j.

We generated test data via the Graph500 unpermuted power law graph generator \cite{bader2006designing}.
We chose the generator because power law distributions well model properties of real world data such as skew \cite{gadepally2015using}.
Generated matrices range from $2^{10}$ rows (scale 10) to $2^{19}$ rows (scale 19),
each with roughly 16 nonzero entries per row.
Multiplying the largest matrices formed close to $2^{33}$ ($= 8 \times 10^9$) partial products.

We used the same Amazon EC2 experiment environment as Section \ref{sExpExpr},
except with 8 workers instead of 4.
Each worker allocated 3 GB of memory to YARN and 3 GB to Accumulo.
The 8-worker environment is well-suited to gauging inter-node parallelism;
intra-node parallelism, however, was limited by the small number of vCPUs (2) per machine.
% we expect greater per-node performance on nodes with more cores and perhaps more disks.
% On the other hand, the modest machine profiles limited cluster cost to \$38 per day.

\begin{figure}[t]
\centering
\includegraphics[width=1\linewidth,clip,trim={0 2px 0 0}]{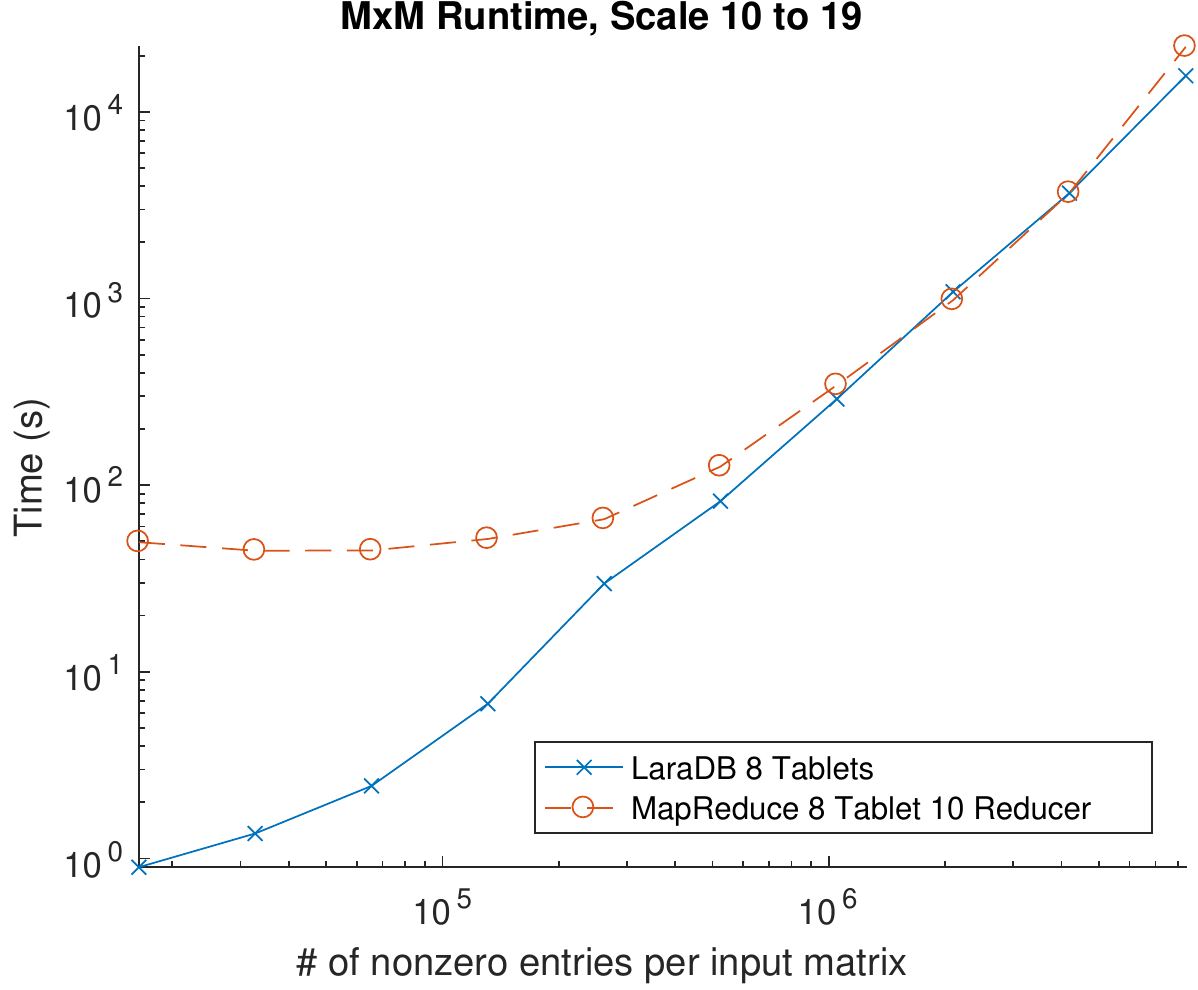}
\caption{8-worker $A^\tr B$ experiment runtime as problem size increases.
\laradb{} dominates at smaller sizes, while 
\laradb{} and MapReduce converge at larger sizes.}
\label{fTime1}
\vspace{-1.5em}
\end{figure}

Figure~\ref{fTime1} plots MxM runtime as problem size increases.
Graphulo dominates MapReduce at smaller problem sizes.
This is due to the large startup cost that MapReduce programs are infamous for;
the YARN scheduler takes roughly 30s to start any task
as a result of job submission, container allocation, jar copying,
and other cold start overheads.

\laradb{}, on the other hand, has a warm start since it runs inside the already-running Accumulo tablet servers.
These tablet servers have a standing thread pool ready to service scan requests as soon as they receive a remote procedure call.
% The threads start the MxM computation right away.
We conclude that \laradb{} is much better suited to interactive and small-scale computation,
such as analytics on a subset of data extracted from an Accumulo table.

At larger problem sizes, \laradb{} and MapReduce converge in performance.
The convergence meets our expectations because the two libraries run similar code
in a similar pattern of parallelism over the same data partitioning.
Their execution environment, JVMs over Hadoop,
is also similar given sufficient time to amortize YARN's startup cost.

We conclude that our \laradb{} implementation is competitive with at least one major RA/LA system at scale.
We take this as initial evidence that systems built atop the \lara{} algebra 
can and do have strong performance.

\section{Conclusion}

Linear algebra (LA) and relational algebra (RA) are, in a sense, two sides of the same coin.
We offer \lara{} as that coin, expressive enough to subsume LA and RA yet with more structure 
than MapReduce that in turn affords greater reasoning.
Lowering \lara{} to a physical algebra brings this reasoning to the domain of partitioned sorted maps, 
a broad abstraction that encompasses LA, RA, and key-value systems 
including the \laradb{} implementation on Accumulo.
% , all while capturing a number of lower-level optimizations.

Our experiments demonstrate that (1) \lara{} expresses high and low-level optimizations that make a difference in the execution of real-world tasks,
and (2) that the \laradb{} implementation outperforms an existing data processing system vastly at small scale and competitively at large scale.

In the future, we aim to use \lara{} as a conduit for studying and computationally exploiting the relationship between LA and RA.
A database optimizer is an ideal place to realize the benefits of this study for joint linear-relational analytics.

\section*{Acknowledgments}
This material is partially supported by NSF Graduate Research Fellowship DGE-1256082.
Thanks to David Maier, Jeremy Kepner, and Tim Mattson for their enthusiasm and comments.
% Opinions, findings, and
% conclusions or recommendations expressed in this material are those of the author and do not
% necessarily reflect the views of the National Science Foundation.

% \renewcommand*{\bibfont}{\footnotesize}
\bibliographystyle{abbrv}
{\small 
\bibliography{header,refs}
}
%\bibliographystyle{abbrv}
%\printbibliography{}
%
% ACM needs 'a single self-contained file'!
%
%\balancecolumns % GM June 2007
\end{document}